\documentclass[a4paper,11pt]{article}
\pdfoutput=1 

\usepackage{jcappub} 
\usepackage[T1]{fontenc} 
\usepackage{subcaption}
\usepackage{pgfplots}
\usepackage{hyperref}
\usepackage{amsmath}
\usepackage{multirow}
\usepackage{placeins}
\usepackage{gensymb}

\newcommand*{\doi}[1]{\href{http://dx.doi.org/#1}{doi: #1}}
\DeclareMathOperator*{\aggfn}{\square}
\title{Application of Graph Networks to a wide-field Water-Cherenkov-based Gamma-Ray Observatory}

\author[a]{J. Glombitza,}
\author[a]{M. Schneider,}
\author[a]{F. Leitl,}
\author[a]{S. Funk,}
\author[a]{and C. van Eldik}

\affiliation[a]{Friedrich-Alexander-Universit\"at Erlangen-N\"urnberg, Erlangen Centre for Astroparticle Physics, Nikolaus-Fiebiger-Str. 2, 91058 Erlangen, Germany}

\emailAdd{jonas.glombitza@fau.de}
\emailAdd{martin.friedrich.schneider@fau.de}
\emailAdd{franziska.leitl@fau.de}

\abstract{
With their wide field of view and high duty cycle, water-Cherenkov-based observatories are integral to studying the very high-energy gamma-ray sky. For gamma-ray observations, precise event reconstruction and highly effective background rejection are crucial and have been continuously improving in recent years. In this work, we investigate the application of graph neural networks (GNNs) to background rejection and energy reconstruction and benchmark their performance against state-of-the-art methods.
In our simulation study, we find that GNNs outperform hand-designed classification algorithms and observables in background rejection and find an improved energy resolution compared to template-based methods.
}


\begin{document}
\maketitle
\flushbottom

\section{Introduction}
\label{sec:intro}
In the last decades, ground-based gamma-ray observatories such as WHIPPLE~\cite{KILDEA2007182}, HESS~\cite{hess_crab}, MAGIC~\cite{ALEKSIC2012435} and VERITAS~\cite{HOLDER2006391} opened a new window to the non-thermal universe at very high energies.
With these observatories, the search for the sources of cosmic rays --- a mystery unsolved for more than 100 years --- as well as astronomy of cosmic phenomena at the highest energies can be performed. 
To survey the gamma-ray sky at very-high energies, search for cosmic particle accelerators, and study diffuse gamma-ray emission within our galaxy, a large field-of-view observatory in the southern hemisphere is required to complement the future Cherenkov Telescope Array (CTA)~\cite{CTA}. 
Since 2019, the Southern Wide-field Gamma-ray Observatory~\cite{abreu2019southernwidefieldgammarayobservatory} (SWGO) collaboration has been working on a next-generation gamma-ray observatory based on water-Cherenkov detectors (WCDs), which has been pioneered by Milagro~\cite{milagro_Atkins_2003, milagrito_ATKINS2000478}, and is also used in the currently operating HAWC~\cite{hawc_Abeysekara_2023} and LHAASO~\cite{lhaasocollaboration2021performance} observatories located in the northern hemisphere.

To maximize the sensitivity of any gamma-ray observatory, a powerful rejection of the dominating cosmic-ray background and precise reconstruction of the air shower properties induced by the impinging gamma ray is crucial.
In the last decades, the reconstruction algorithms~\cite{smith2015_reco_hawc} have been continuously improved, e.g., using template-based methods~\cite{template_Parsons_2014, templates_vikas} and tree-based machine learning techniques applied to $\gamma$\;/\;hadron separation at Imaging Air Cherenkov Telescopes (IACTs)~\cite{OHM2009383, boosted_decision_trees_veritas_Krause_2017, random_forest_magic_Albert_2008}.

The latest progress in machine learning algorithms --- based on deep neural networks (DNNs) --- called \emph{deep learning}~\cite{deeplearning}, provides novel techniques for enhancing reconstruction algorithms in the physical sciences~\cite{deeplearning, dlfpr}.
Even the first application of neural networks to gamma-ray astronomy dates back to the 90s~\cite{hegra_geiger_neural_netwok_WESTERHOFF1995119} and early 2000s~\cite{neural_network_magic_BOINEE_2006}; to this time, they have not been capable due to their limited capacity to exploit the patterns in the recorded data in full detail.
This has changed with \emph{deep} networks that are able to analyze even tiny patterns in data with unprecedented precision.
In turn, using DNNs, the event reconstruction capabilities for cosmic-ray research~\cite{ERDMANN201846, xmax_wcd, pao_muon_dnn}, neutrino~\cite{abbasi_convolutional_2021,km3net_dl}-, and gamma astronomy~\cite{Bom:2021abs, assunccao2019automatic, 4PMTs_NCA, WCD4PMTs, Watson:2023vx} could be significantly improved, leading to new insights into the universe at the highest energies~\cite{gal_plane_2023, thepierreaugercollaboration2024inference, thepierreaugercollaboration2024measurement}.

In this work, we investigate the application of DNNs to background rejection and energy reconstruction for a baseline design currently developed for SWGO.
The graded layout considered in this work comprises around 4600 detector stations, and the number of triggered stations per event varies from tens to thousands but is small on average.
Thus, we decided to use graph convolutional neural networks (GNNs)~\cite{geo_dl} to analyze the sparse signal patterns.
The use of GNNs enables both the efficient analysis of signal patterns of different sizes and the exploitation of the advantages of convolutional neural networks, a central driving force in the recent success of deep learning~\cite{deeplearning}.
With our GNN approach, we find significant improvements in hadronic background rejection compared to previous, hand-designed observables and classification algorithms and a promising energy reconstruction slightly surpassing the performance of currently state-of-the-art template-based reconstruction chains~\cite{templates_vikas}.

\section{Simulation of a WCD observatory in the southern hemisphere\label{sec:data}}
To prepare the data for our analysis, we use simulated events obtained from Monte-Carlo (MC) simulations for a baseline design currently developed within the SWGO collaboration.
Similar to HAWC, SWGO uses a combination of standard simulation packages in use in the gamma-ray astronomy community, namely CORSIKA~\cite{heck_1998} for air shower simulations and GEANT4~\cite{AGOSTINELLI2003250} for the interaction of shower particles with the detector.

\subsection{Simulated Detector Design}
\begin{figure}[t!]
    \centering
    \subfloat[Detector layout]{\includegraphics[width=0.475\textwidth]{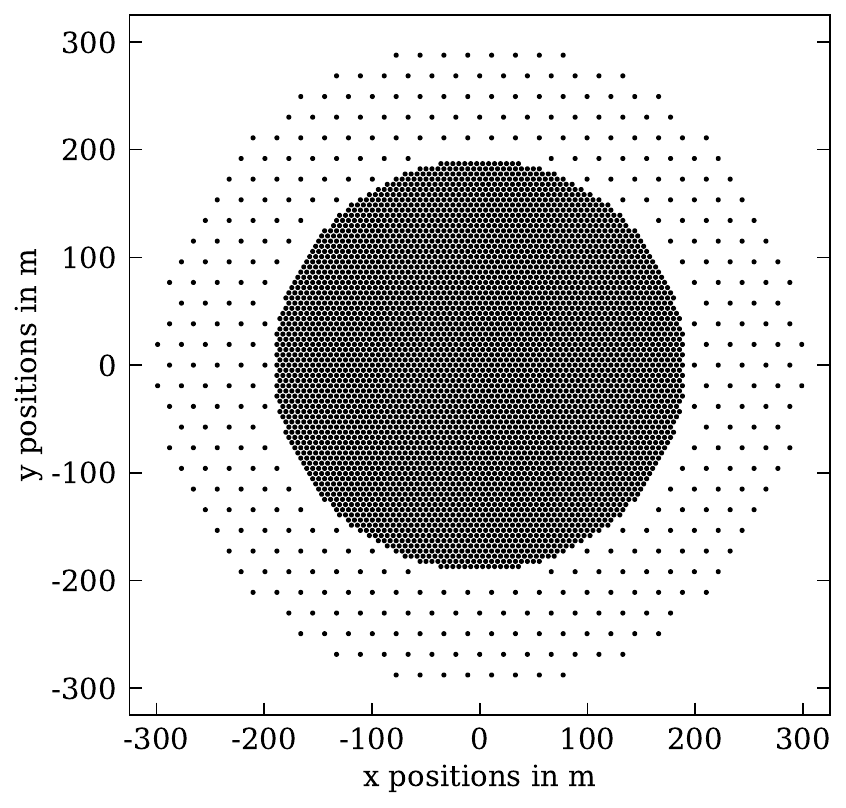}}
    \subfloat[Tank design]{\raisebox{0.3\height}{\includegraphics[width=0.35\textwidth]{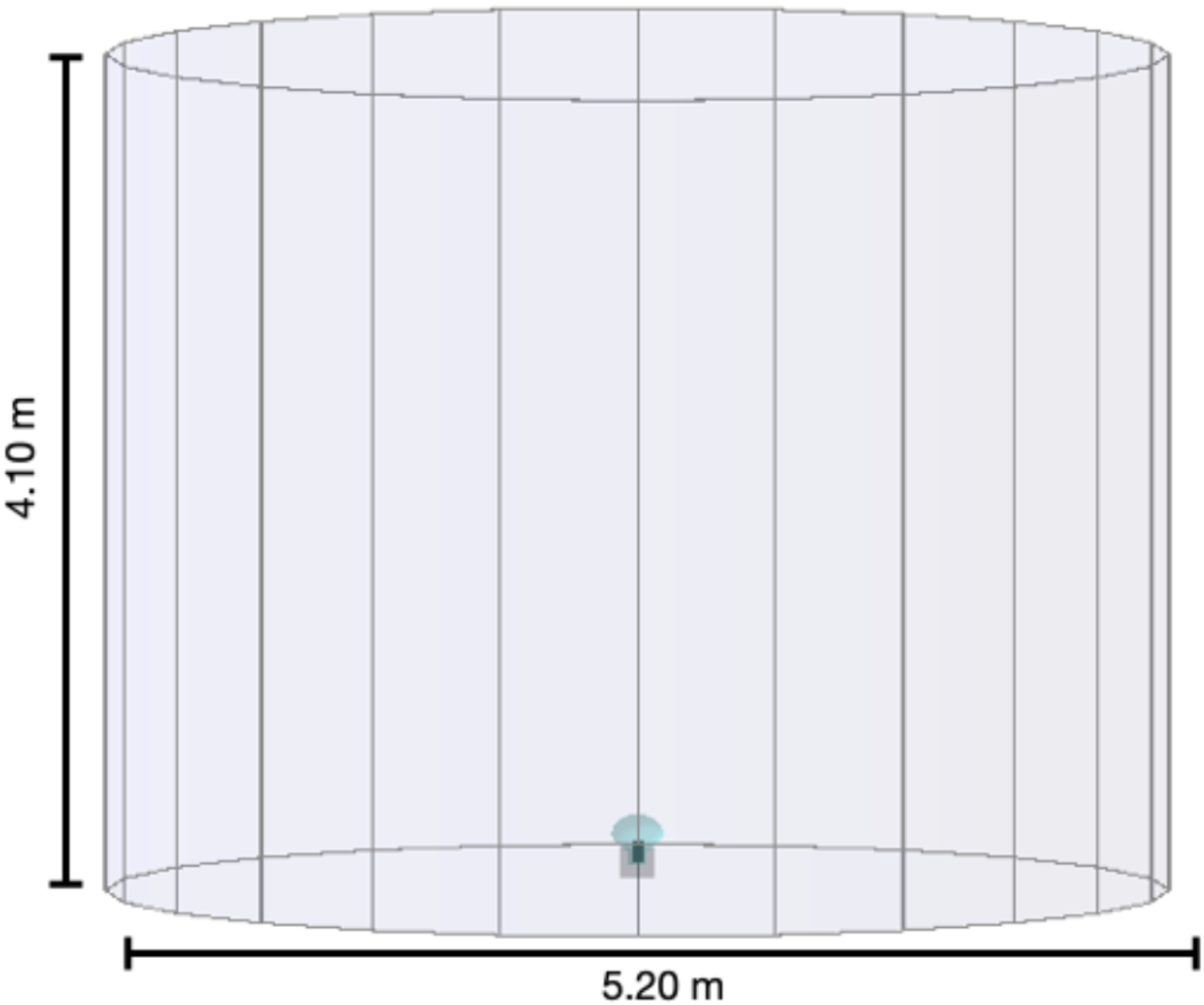}}}\hfill
\caption{Detector configuration used in this work, currently investigated as a candidate design for SWGO.
Left: Station layout featuring two zones with fill factors of 80\% (central zone) and 5\% (outer zone).
Right: Used water-Cherenkov tank design. The tank contains a single upward-looking 10-inch-PMT (Credit: SWGO Collaboration~\cite{ruben2023southernwidefieldgammarayobservatory}, reproduced with permission). 
} 
\label{fig:point_cloud}
\end{figure}

The stations in the considered design feature a large tank with a diameter of 5.20~m and a height of 4.10~m with an upward-looking 10-inch PMT located at the center of the tank floor, as shown in the right of Figure~\ref{fig:point_cloud}.
As such, it is very similar to the LHAASO and HAWC station designs.
The studied layout comprises around 4600 detector units covering an area of roughly $280{,}000\,\mathrm{m}^2$ and features a graded layout with two zones of different fill factors.
Whereas in the central zone of a 188~m radius, the tanks are densely packed, reaching a fill factor of $80\%$ to increase the sensitivity at low energies, the outer zone features a fill factor of $5\%$ to increase the effective area at low cost.

We simulated this detector layout with the established HAWCsim package of HAWC~\cite{Abeysekara_2017}, which was modified for SWGO~\cite{Schoorlemmer:2021lle}.
This enables a realistic simulation of the complete detector response, which includes photomultiplier tube (PMT) noise, but still lacks cosmic ray-induced noise hits.
The simulation set amounts to roughly 440,000 proton events and 370,000 gamma-ray events with a threshold of $>25$ detector hits.
This is slightly lower than the trigger threshold of 30 hits\footnote{A first approximation of the threshold realizable for the examined design assuming an available bandwidth of 2 Gb/s.} investigated in the following, to expand generalization capacities of the trained network at the edges of the phase space.
The data is split 70\%/10\%/20\% into training, validation, and test data sets.
The events cover an energy range between 31.6\,GeV and 1\,PeV following a spectral index of $-2$, and zenith angles up to 65$^\circ$.
We use the same spectral index for protons and gamma rays to make the classifier response independent of the energy spectrum, which could lead to increased systematic uncertainties when, in reality, the gamma source spectrum differs from the one used during training.

\subsection{Data preparation and preprocessing\label{sec:prepro}}
To exploit the air shower footprint for event reconstruction and $\gamma$\;/\;hadron separation, we use the arrival time distribution of the particles detected at ground level.
Since exploring the detailed temporal structure of the PMT signal is computationally quite demanding, we make use of integrated signal, i.e., charge, and do not use the signal trace directly.
Thus, we use the $x,y$ positions of the tanks, the arrival times $t$ of the first Cherenkov photon hitting the PMT of each station, as well as the measured charge of each PMT as input into our DNN.
Each point in this point cloud resembles the measurements of the PMT of a single tank.
The location of each point is given by the $x$ and $y$ coordinates of the tanks, and the arrival time $t$ and signal charge $q$ of the PMT form additional features of each point. 
Thus, each point holds $2 + 2$ features for each single PMT.

\subsubsection{Graph construction}
\begin{figure}[t!]
    \centering
    \subfloat[Time]{\includegraphics[width=0.5\textwidth]{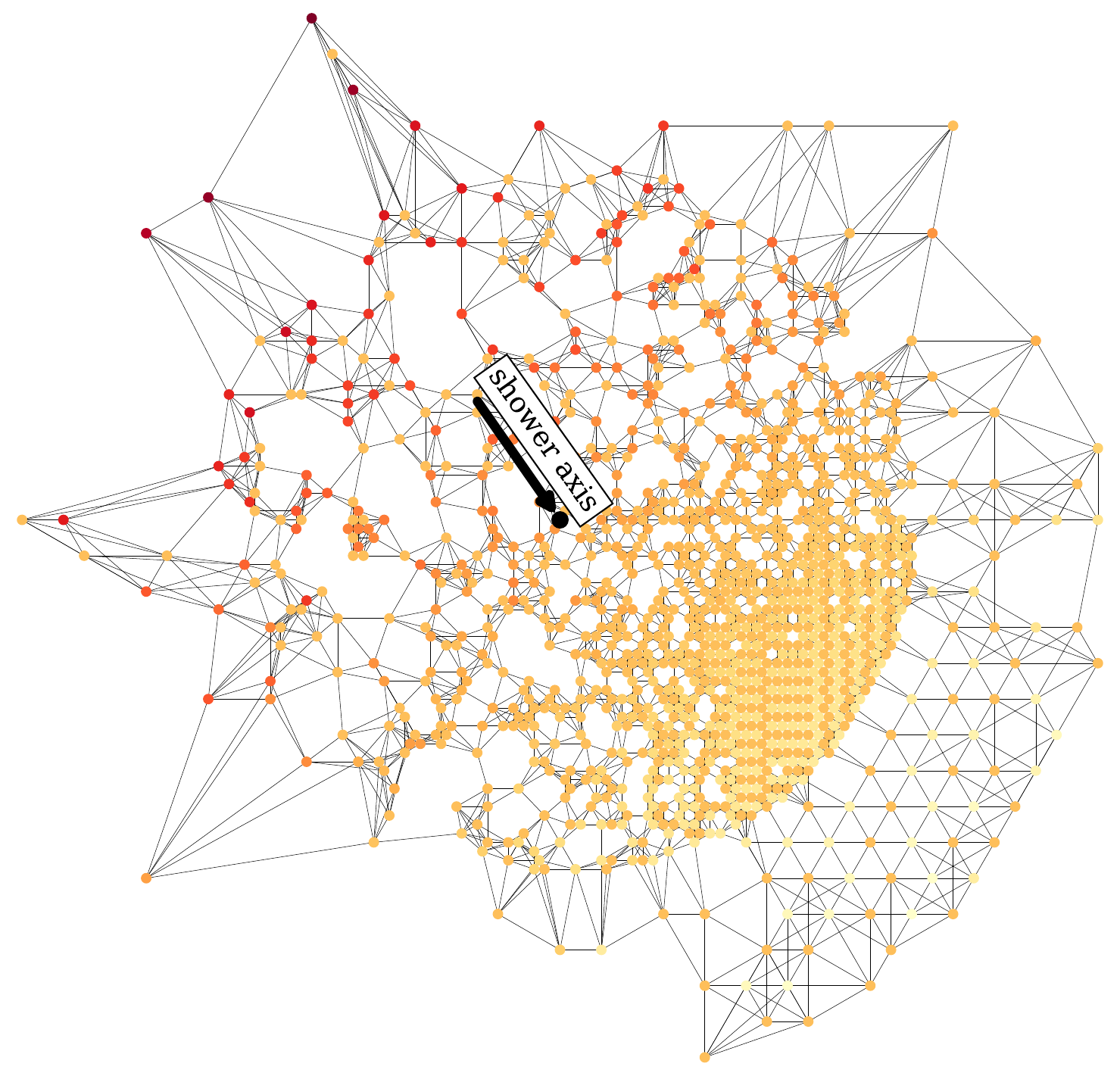}}
    \subfloat[Charge]{\includegraphics[width=0.5
\textwidth]{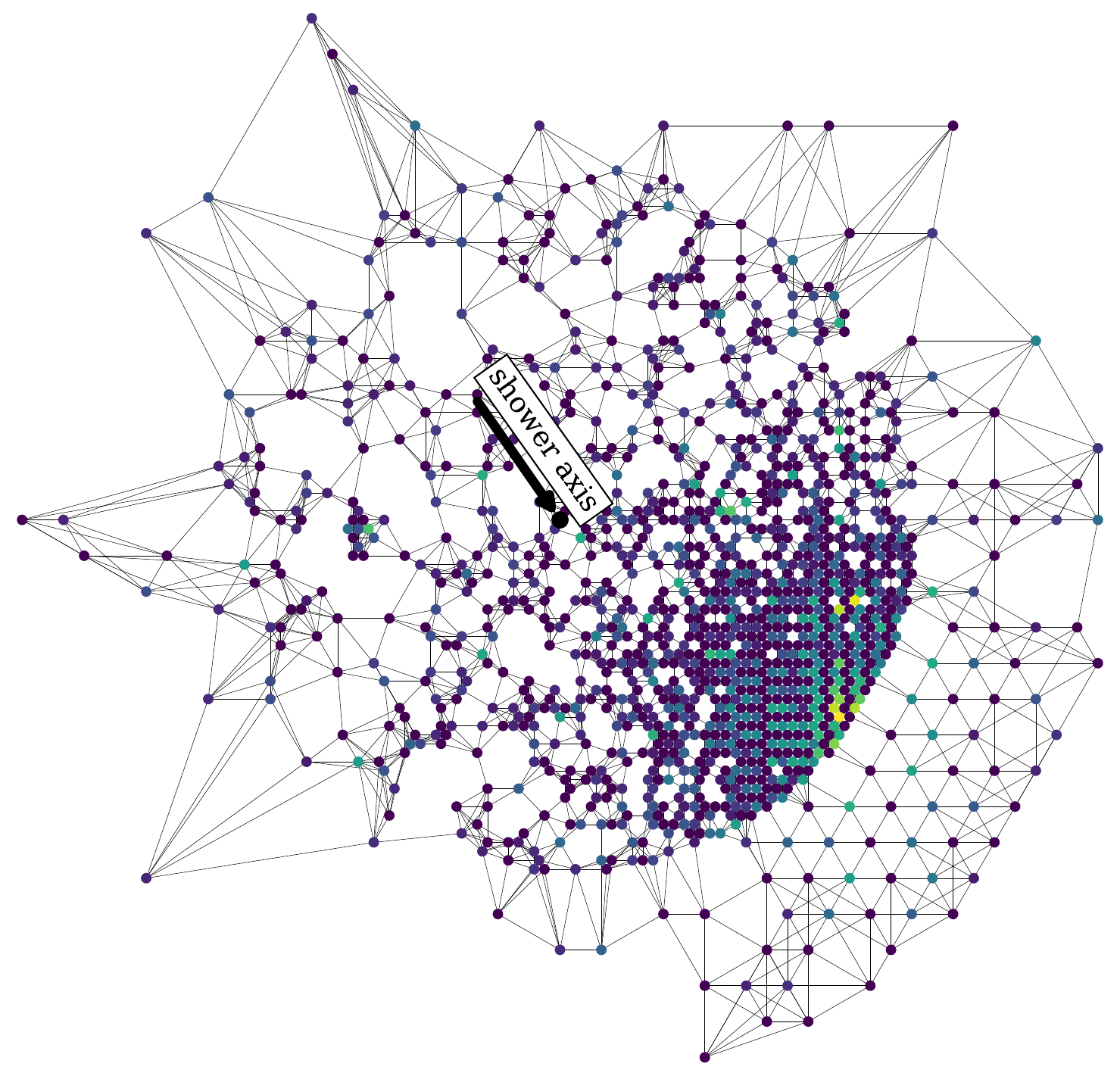}}\hfill
\caption{Graph representation of a simulated gamma-ray event with a true energy of $18.8\,$TeV and a zenith angle of $39.7^\circ$ obtained after $k$nn clustering. 
Left: Detected arrival time, yellow=early, red=late.
Right: Deposited charge, blue=low, yellow=high.}
\label{fig:example_graph}
\end{figure}

The point cloud is used for the construction of the graph in the following way.
We apply a $k$-next neighbor clustering ($k$nn) in the spatial dimension to define a graph using the $N$ triggered stations of a given event, including self-connections.
Each node in the graph will be connected to itself and its six nearest neighbors, which are determined based on the Euclidean distance between the points. The number of six neighbors is motivated by the triangular grid of the layout of SWGO, as each tank has six direct neighbors. This yields a directed graph with $N$ nodes and $7\times N$ edges.
An example of such a feature graph of an example event is shown in Figure~\ref{fig:example_graph}.
The arrival times (left) and charges (right) are color-encoded.

\subsubsection{Normalization}
To stabilize the GNN training, we perform feature normalization of the positions, PMT charge, and time.
For the timing information z-score normalization~\cite{dlfpr} is used $t^\prime = (t-\mu)/\sigma_\mathrm{std}$ while a logarithmic re-scaling is performed for the charges $q^\prime = \big(\mathrm{log}_{10} \,(1 + q)\big) / \sigma_\mathrm{std}$.
Here, $\mu$ denotes the mean and $\sigma_\mathrm{std}$ the standard deviation of the time and charge\footnote{After applying the transform $\mathrm{log}_{10}(q+1)$.} distribution estimated over all events, respectively.
For the station positions, we also perform z-score normalization.

\section{Graph neural networks for SWGO}
The footprint of a typical gamma-ray shower at several TeV roughly covers an area of hundreds of $\mathrm{m}^2$.
For the currently discussed detector designs for SWGO~\cite{ruben2023southernwidefieldgammarayobservatory}, this translates into tens to thousands of detector stations, forming a set of triggered PMTs at ground level. Image-like data can be precisely and efficiently analyzed with deep learning techniques using convolutional neural networks (CNNs)~\cite{deeplearning, dlfpr}.
In CNNs, small filters with adaptive parameters exploit the image information by enforcing translational invariance and setting a prior on local correlations.
However, the use of a filter with a fixed size in CNNs requires the analyzed data to be distributed on a homogeneous grid, which should have low sparsity to ensure efficient usage.
The strong variation in the number of triggered stations and the graded layout makes using CNNs for analyzing the SWGO raw data unfeasible since either the algorithm training is inefficient or information needs to be dropped~\cite{assunccao2019automatic}.
In the scope of this work, we model an SWGO event as a signal graph consisting of nodes and edges, where the nodes are formed by the triggered stations holding the information of the station's PMTs as node features $\mathbf{x}_i = (x, y, q,t)^T$, and the edges connect stations with local proximity.\\

\subsection{Network design}
We make use of EdgeConvolutions~\cite{wang2019dynamic} in this work that were already successfully applied to various physics challenges~\cite{Qu_2020, Abbasi_2022_graph, Bister_2021, Glombitza_2023}.
The EdgeConvolution can be divided into graph construction, convolution, and aggregation.
After the graph generation (see Sec. \ref{sec:prepro}), we regard the immediate neighborhood $\mathbf{x}_j$ of each "central" node $\mathbf{x}_i$, which depends on a fixed number of $k$NN next neighbors.
In this study, we use $k=6$. 
The EdgeConvolution of nodes and aggregation of information is defined as:
\begin{equation}
    \mathbf{x}^{\prime}_i = \sigma\Big(\aggfn_{j \in \mathcal{N}(i)} h_{\mathbf{\Theta}}(\mathbf{x}_i \, \Vert \, \mathbf{x}_j - \mathbf{x}_i)\Big).
\end{equation}
Each pair of $(\mathbf{x}_j - \mathbf{x}_i)$, in the neighborhood of each $\mathbf{x}_i$, is convolved leading to a number of $f$ different edge features.
Note that the $\Vert$ indicates that the subtraction is not applied to the central node, i.e., the self-connection.
This convolution is performed by the kernel function $h_{\mathbf{\Theta}}$ with output dimension $f$, to be chosen by the user.
The kernel function combines the information of the feature vector of $\mathbf{x}_i$ with the edge features of its neighboring nodes.
Finally, the $k$ edge features of the $k$ neighbors will be aggregated using an aggregation function $\aggfn_{j \in \mathcal{N}(i)}$ and later passed through an activation function $\sigma$.
Such aggregation functions combine, i.e., aggregate, information from neighbors of a node to produce a summary representation. Possible choices include mean, maximum, or the sum (used in this work), which means that the output remains independent of the order of the neighbors, conserving permutational invariance.
This leads to the feature vector $\mathbf{x}^{\prime}_i$ after the edge convolution, which is located at the position of the central node.
This convolutional operation is performed in parallel for each node $\mathbf{x}_i$ of the graph.

Each EdgeConvolution layer disseminates information in its immediate vicinity, giving local information to the network.
To exploit beyond local and global correlations across the whole graph, we stack multiple graph layers to ensure a large receptive field of view~\cite{dlfpr}.
Furthermore, DynamicEdgeConvolutions~\cite{wang2019dynamic} can be utilized, to extract global features, which perform a graph construction step during training by finding the $k$-closest nodes in the feature space.
Since graph construction causes computational overhead, as it needs to be recalculated for each layer and sample during training, we use DynamicEdgeConvolutions in a limited fashion.

\begin{figure}[]
    \centering
    \includegraphics[width=0.575\textwidth]{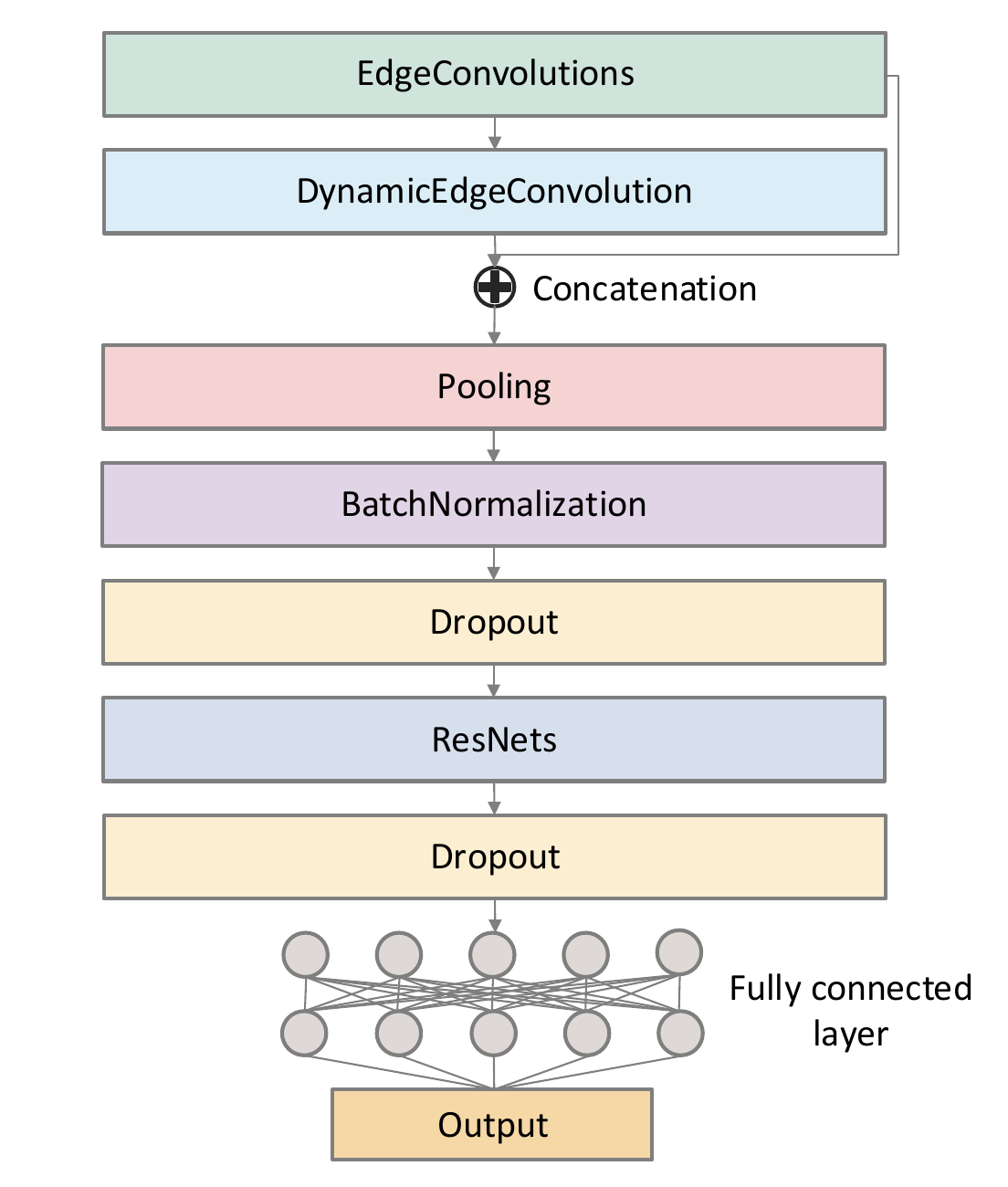}
    \caption{Sketch of the general network structure used for the $\gamma$\;/\;hadron separation and energy reconstruction.
    The network consists of a number of EdgeConvolution blocks, followed by a DynEdgeConvolution layer. After the concatenation of the layers, the result is propagated through batch normalization, dropout, ResNet blocks, and another dropout layer.
    Finally, a small, fully-connected layer is applied, which leads to the final output of the network.}
    \label{fig:network_sketch}
\end{figure}

In this analysis, the general network structure for each task was inspired by the ParticleNet network described in Ref.~\cite{Qu_2020}.
A sketch of the general network structure can be found in Figure~\ref{fig:network_sketch}.
The detailed network structure used for the $\gamma$~/~hadron separation and energy reconstruction can be found in the appendix Sec.~\ref{sec:gnn_details} (see Tab.~\ref{tab:separation_arch} for the $\gamma$~/~hadron separation and Tab.~\ref{tab:energy_reco_arch} for the energy reconstruction).
In general, both networks consist of a number of consecutive EdgeConvolutions to gather local information and a DynamicEdgeConvolution block to extract global information from the air shower footprint.
Depending on the reconstruction task, the number of EdgeConvolution and DynamicEdgeConvolution layers vary, as we found that different architectures exhibit slightly improved GNN performance optimized for the used layout configuration.
Similar to Ref.~\cite{Qu_2020}, the kernel function $h_\mathbf{\Theta}$ of the EdgeConvolution consists of three fully-connected layers where each is being followed by a batch normalization and an activation function, see Tab.~\ref{tab:edge_conv_ffn} for all the details.
Next, a global pooling layer, performing pooling along the node dimension, is used to make the final part of the network independent of the number of triggered stations.
Finally, a set of task-specific fully-connected layers containing the output follows, completing the GNN structure.

\subsection{Training}
We performed extensive hyperparameter tuning for $\gamma$\;/\;hadron separation and energy reconstruction to converge to the final GNN architecture for each task.
The best models for each configuration are chosen based on the lowest observed validation loss.
All networks were implemented using PyTorch Geometric~\cite{pyg}, and trained on single NVIDIA A40 or A100 GPUs, with training times ranging from 8 up to 24 hours depending on the hyperparameter configuration. We used binary cross-entropy for g/h separation and mean-squared-error as the loss function for energy reconstruction. During training, we applied a strategy that involved reducing the learning rate upon reaching a plateau and incorporated early stopping. Thus, the learning rate was lowered if the validation loss did not improve for five epochs, and training was stopped entirely if no improvement was observed after 11 epochs.\\
A hyperparameter search of 70 trainings is performed for each task-specific network, varying different parameters using a random search. 
As optimizer, we used Adam~\cite{kingma_adam_2017}.
We vary parameters that affect the learning process, such as learning rate, weight decay, and decay factor\footnote{The decay factor is used to reduce the learning rate after a certain number of epochs without improvements in the validation loss.}, dropout~\cite{JMLR:v15:srivastava14a}, and if batch normalization~\cite{ioffe2015batchnormalizationacceleratingdeep} is used, as well as parameters like the number of graph convolutional layers, the size of $k$ in the clustering step of DynamicEdgeConvolution, the number of kernel features $f$, and the number of ResNet blocks\footnote{ResNet blocks refer to a stack of two fully-connected layers with a shortcut connection that is utilized to perform an identity mapping by simply adding the inputs to the outputs of the stacked layers.
This learning of residual functions has been proven to ease the training of DNNs~\cite{he2015deep}.} (see Tab.~\ref{tab:resnet_arch} for the ResNet architecture).
Note that dropout is applied once after the graph pooling and once at the end of the output of the last ResNet layer.
A detailed table of the final parameters of the search is shown in Tab.~\ref{tab:edge_conv_arch}.

For the energy reconstruction, we apply a minor modification of each convolution block by using a residual connection in contrast to the $\gamma$\;/\;hadron separation model.
As required by ResNet, the input of each convolution block is projected via a fully connected layer prior to the addition of the output of the convolutional layer to ensure similar dimensions.
Finally, a ReLU activation is applied.
The energy of each predicted event is given in $\log_{10}\left(E/ \rm{GeV}\right)$.
More information on the architecture and the validation losses can be found in the Appendix~\ref{sec:gnn_details}.

\begin{figure}[t!]
    \centering
    \includegraphics[width=0.95\textwidth]{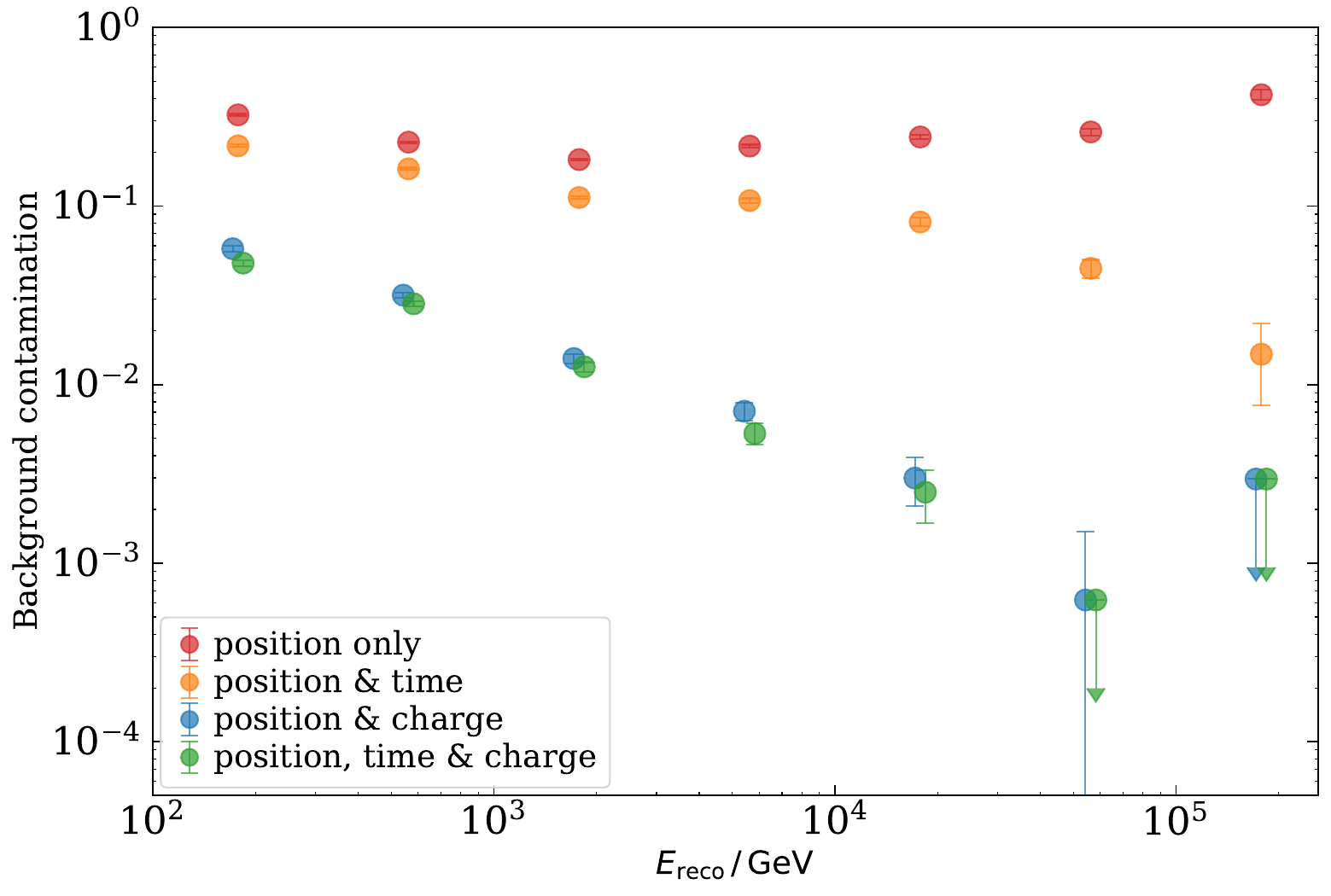}
    \caption{Comparison of the background rejection performance, when training the GNN using different inputs. 
    The residual background contamination using all available features is shown in green. 
    Orange and blue show the network performance with positions and time or positions and charge used as input features. 
    In red, the performance of the network utilizing only positional information is shown. Error bars denote statistical uncertainties following Ref.~\cite{ullrich2012treatmenterrorsefficiencycalculations}}
    \label{fig:gnn_feature_comparison}
\end{figure}

\section{$\boldsymbol{\gamma}$ / hadron separation using GNNs}
In the following section, we investigate the performance of the GNN in terms of $\gamma$\;/\;hadron separation.
Our trained GNN model output is a score that indicates the likelihood of the event to be a gamma ray (output close to 1) or a proton (output close to zero)\footnote{Two example score distributions (protons = blue, gamma=red) are shown in Figure~\ref{fig:score_plot}.}. 

\subsection{Performance metrics}
For estimating the performance of the trained GNN classifier, we study the false positive rate $\epsilon_p$ (fraction of protons relative to the simulated number of protons misclassified as gamma rays), in the following referred to as \emph{background contamination}, as a function of reconstructed energy~\cite{LHLatDistFit_PoS2023}.
We study the energy dependence of the background contamination for a fixed gamma-ray efficiency $\epsilon_\gamma$, i.e., the surviving fractions of gamma rays, which is usually set to be $\approx80\%$~\cite{albert2024performancehawcobservatorytev}.

We also make use of the \emph{quality factor}, which combines proton contamination and gamma-ray efficiency and is defined via:
\begin{equation}
    Q = \frac{\epsilon_\gamma}{\sqrt{\epsilon_p}}.
\end{equation}
Assuming that each bin contains a sufficiently large number of events, $Q$ is proportional to the discovery Poisson significance improvement of a source.
By changing the score threshold of the classifier, i.e., the score at which an event is tagged as a gamma ray, the quality factor will be maximized\footnote{An example is shown in Figure~\ref{fig:score_plot}, alongside the quality factor as a function of the threshold on the GNN output for the low energy and high energy regime.} to optimize sensitivity~\cite{milagro_Atkins_2003}.

\begin{figure}[t!]
    \centering
    \subfloat[$2.5 \leq \log_{10}\,(E_\mathrm{reco}/\mathrm{GeV}) < 3.0$]{\includegraphics[width=0.49\textwidth]{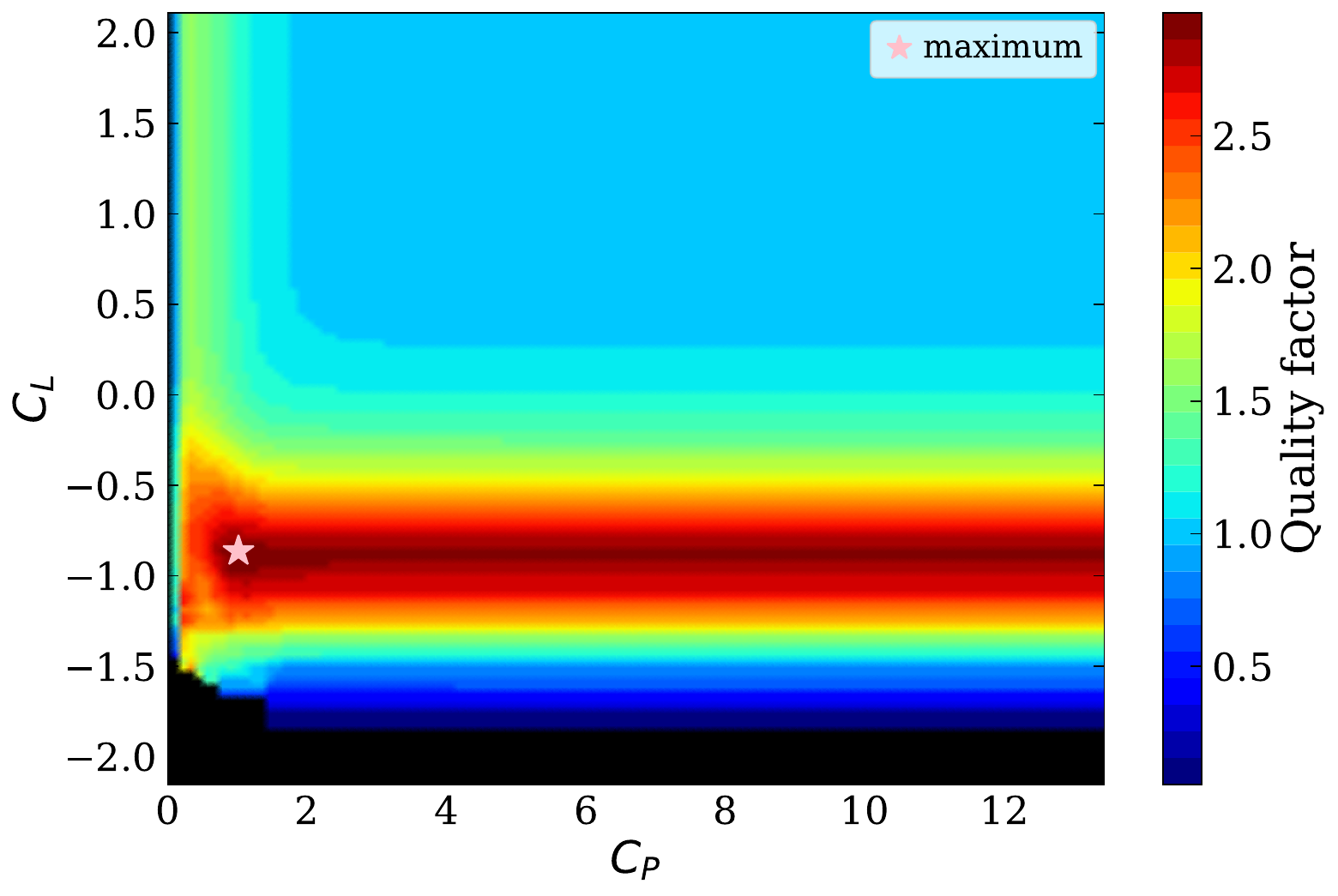}}
    \subfloat[$4.5 \leq \log_{10}\,(E_\mathrm{reco}/\mathrm{GeV}) < 5.0$]{\includegraphics[width=0.49
\textwidth]{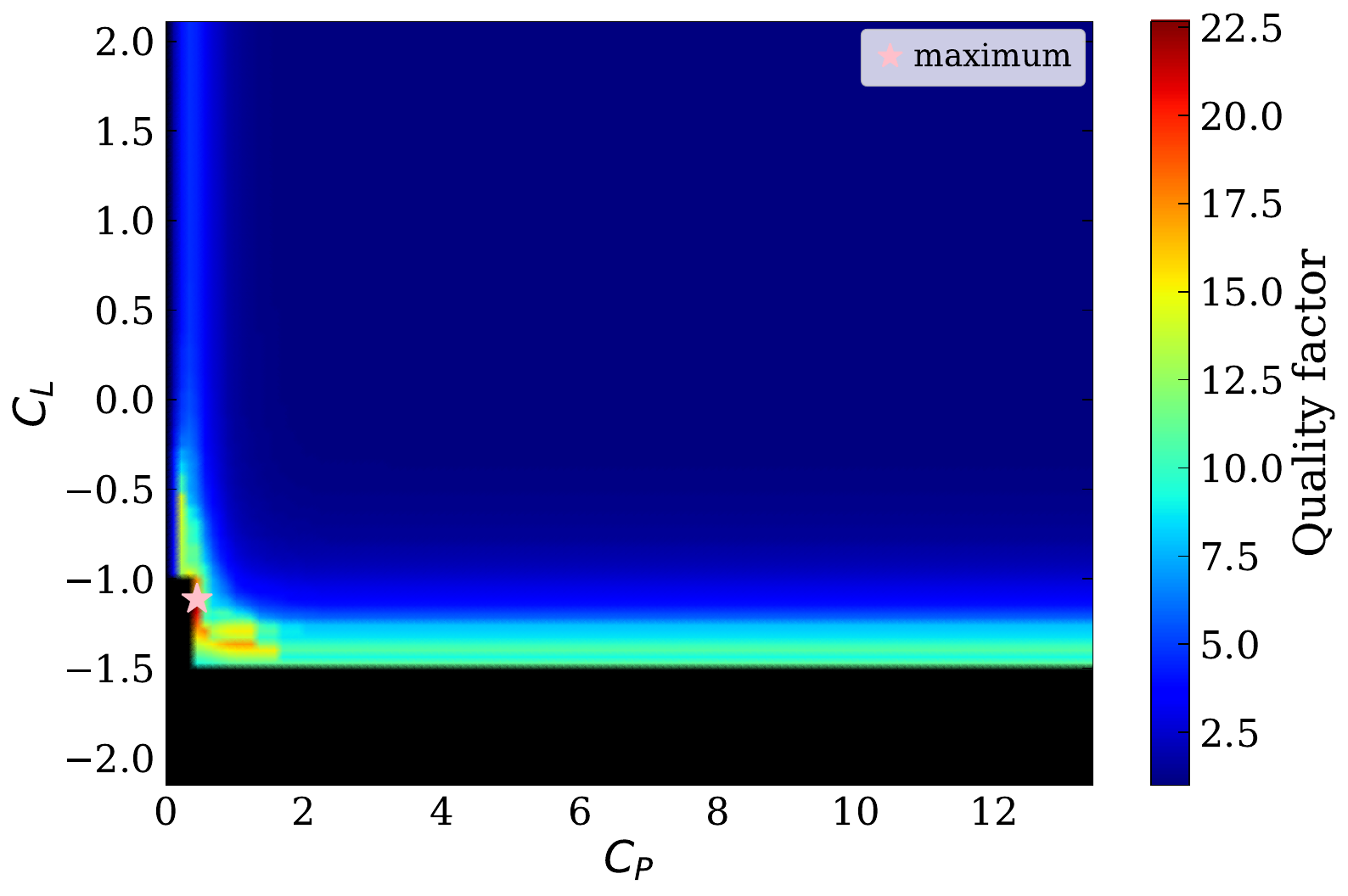}}\hfill
\caption{The quality factor shown as a function of a cut on PINCness ($C_P$) and LIC ($C_L$) for two different energy bins. The maximum of the quality factor is indicated by a pink star. The thresholds $C_P$ and $C_L$ at this point provide the optimal $\gamma$\;/\;hadron separation in terms of maximizing the sensitivity as estimated by the quality factor and are used in the following.\label{fig:box_cut_plot}}
\end{figure}

\subsection{Separation performance using different features}
As a first step, we study the impact of different inputs: Charge, arrival time, and the positions of triggered stations, i.e., the signal pattern induced by the shower footprint, for the network architecture.  
To study the $\gamma$\;/\;hadron separation performance using only positional information, the network is trained using two positional features $x$ and $y$ and a feature representing the PMT signal, which was set to random uniform noise, with unit variance and zero mean.
Additionally, the performance of adding either the timing or charge information is investigated by using this information as a feature in place of the noise.
The performance is shown in Figure~\ref{fig:gnn_feature_comparison} using the events of the test data set.
Error bars denote statistical uncertainties and are estimated following Ref.~\cite{ullrich2012treatmenterrorsefficiencycalculations}.
Utilizing only positional information (red markers), i.e., the position of triggered stations, the GNN has only weak $\gamma$\;/\;hadron separation capabilities.
Adding time as a feature improves $\gamma$\;/\;hadron separation performance, with a stronger gain at higher energies.
Using only charge instead of time information yields excellent separation performance, which is not surprising, given that mostly signal-based classifiers~\cite{milagro_Atkins_2003, Abeysekara_2017, Hampel-Arias:20164i, ruben_az_fluc_2022, albert2024performancehawcobservatorytev} primarily focusing on detecting signal fluctuations in the shower footprint, i.e., hadronic sub showers, were used for $\gamma$\;/\;hadron separation in the past.
The best result is obtained by using both timing and charge information as input, which slightly improves the performance over the whole energy range, resulting in an overall background contamination ranging from 5\% at low energies to below 0.1\% at the highest energies.

\begin{figure}[t!]
    \centering
    \includegraphics[width=0.9\textwidth]{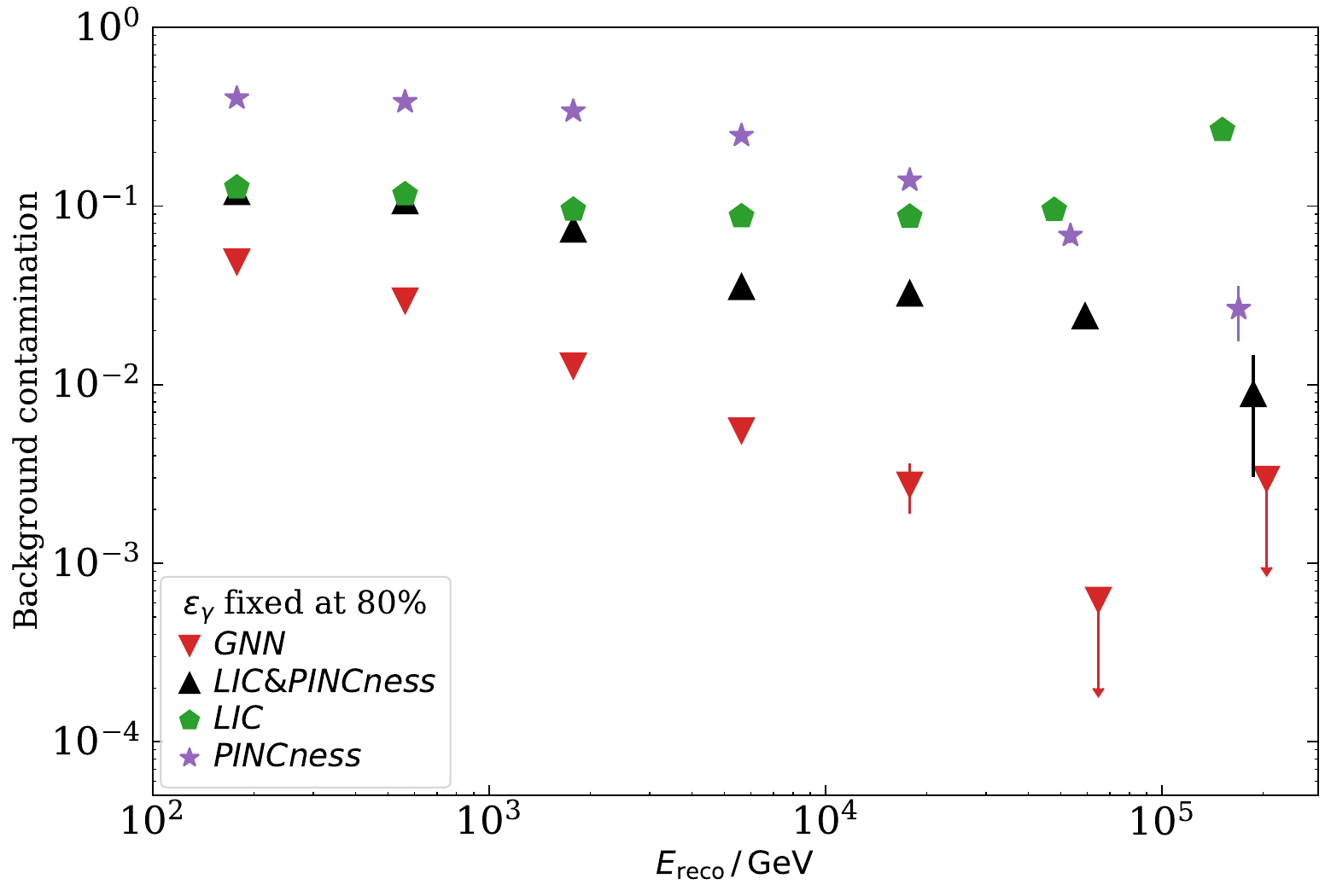}
    \caption{$\gamma$\;/\;hadron separation performance comparing the GNN (red triangles) to PINCness (purple stars), LIC (green pentagons), and their combination (black triangles). The gamma-ray efficiency was fixed to 80\% to enable a fair comparison. The data points in the last two bins shifted horizontally to improve readability. Error bars denote statistical uncertainties. Upper limits are estimated by assuming a single background event and are depicted by downward-facing arrows.}
    \label{fig:contamination_comparison_fixed_gammaeff}
\end{figure}

\subsection{Comparison to established classification observables}
In the following subsection, we compare the GNN performance to two established observables that have proven their effectiveness and are currently used within HAWC~\cite{albert2024performancehawcobservatorytev}.

\subsubsection{Definition of baselines observables}
The $\gamma$\;/\;hadron separation utilized in HAWC relies on these two distinct parameters, called LIC and PINCness, to distinguish between cosmic-ray and gamma-ray events.
The $\mathrm{LIC}$ parameter is the logarithm of the inverse of the so-called compactness~\cite{milagro_Atkins_2003}, originally developed for Milagro:
\begin{equation}
    \mathrm{LIC} = \log_{10}\left(\frac{CxPE_{40}}{n_{\mathrm{hit}}}\right).
\end{equation}
Here, $CxPE_{40}$ denotes the largest charge measured in a PMT at least 40\,m away from the shower core, and $n_{\mathrm{hit}}$ is the number of tanks hit in the event (our example design features only a single PMT per tank).

The second parameter, PINCness~\cite{Abeysekara_2017}, makes use of the lateral distribution function to separate gamma rays from cosmic rays.
It is defined as 
\begin{equation}
\mathrm{PINCness} = \frac{1}{N}\sum^N_{i=0} \frac{\left( \log_{10}(q_i) - \langle \log_{10}(q_i) \rangle \right)^2}{\sigma^2},
\end{equation}
where $q_i$ is the charge of the $i$-th PMT that was triggered in the event and $\sigma$ is the uncertainty on the charge. 
It is calculated by fitting Gaussian distributions to the logarithm of average charges for different rings centered on the shower core and then fitting the best-fit values to a quadratic function.

Given that simple box cuts on these two parameters are widely used to combine both parameters for $\gamma$\;/\;hadron separation, we optimize them in the same way as Ref.~\cite{Alfaro_2022} to define a baseline to compare to.
We find the optimal cuts for the LIC and PINCness parameters by optimizing the quality factor for every bin in our dataset, which consists of seven logarithmic reconstructed energy bins from $10^{2.0}$\,GeV to
$10^{5.5}$\,GeV with a step size of half a decade. Figure~\ref{fig:box_cut_plot} shows the best cuts obtained for one low-energy bin and one high-energy bin.

\begin{figure}[t!]
    \centering
    \subfloat[$2.5 \leq \log_{10}\,(E_\mathrm{reco}/\mathrm{GeV}) < 3.0$]{\includegraphics[width=0.49\textwidth]{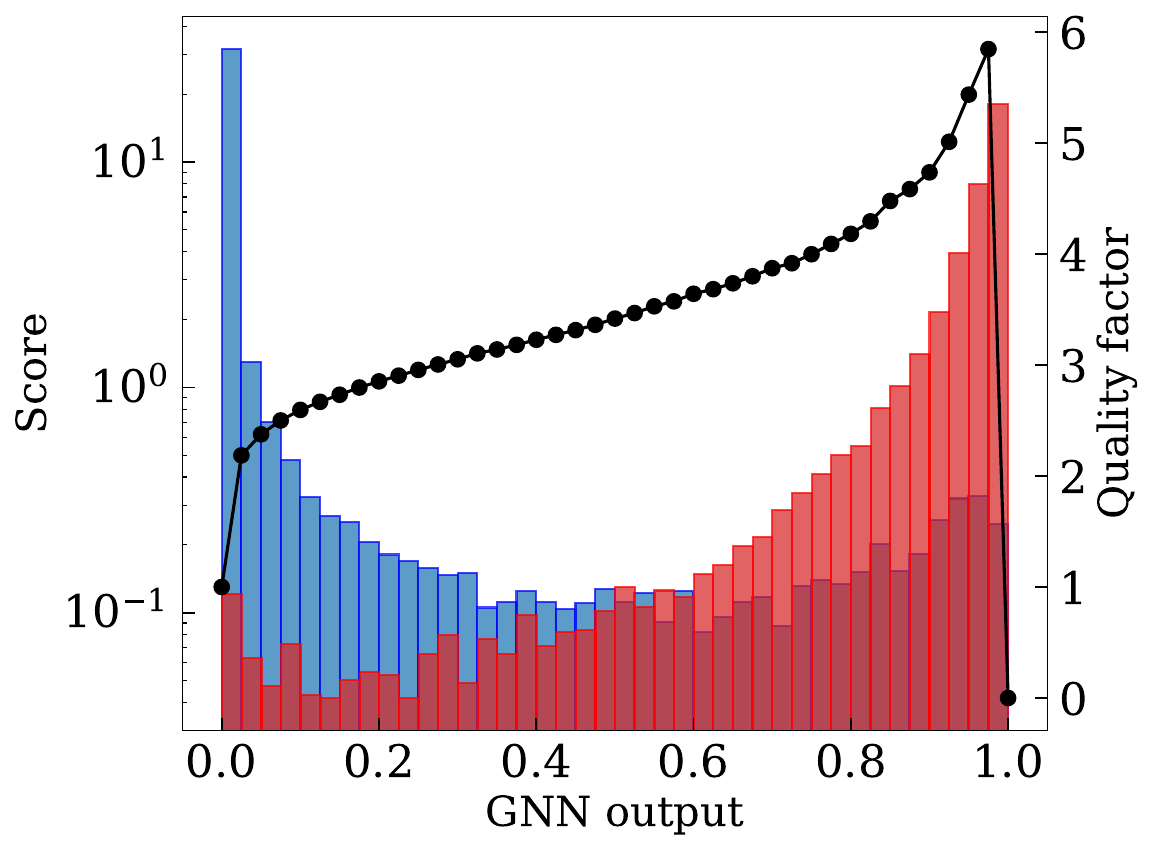}}
    \subfloat[$4.5 \leq \log_{10}\,(E_\mathrm{reco}/\mathrm{GeV}) < 5.0$]{\includegraphics[width=0.49
\textwidth]{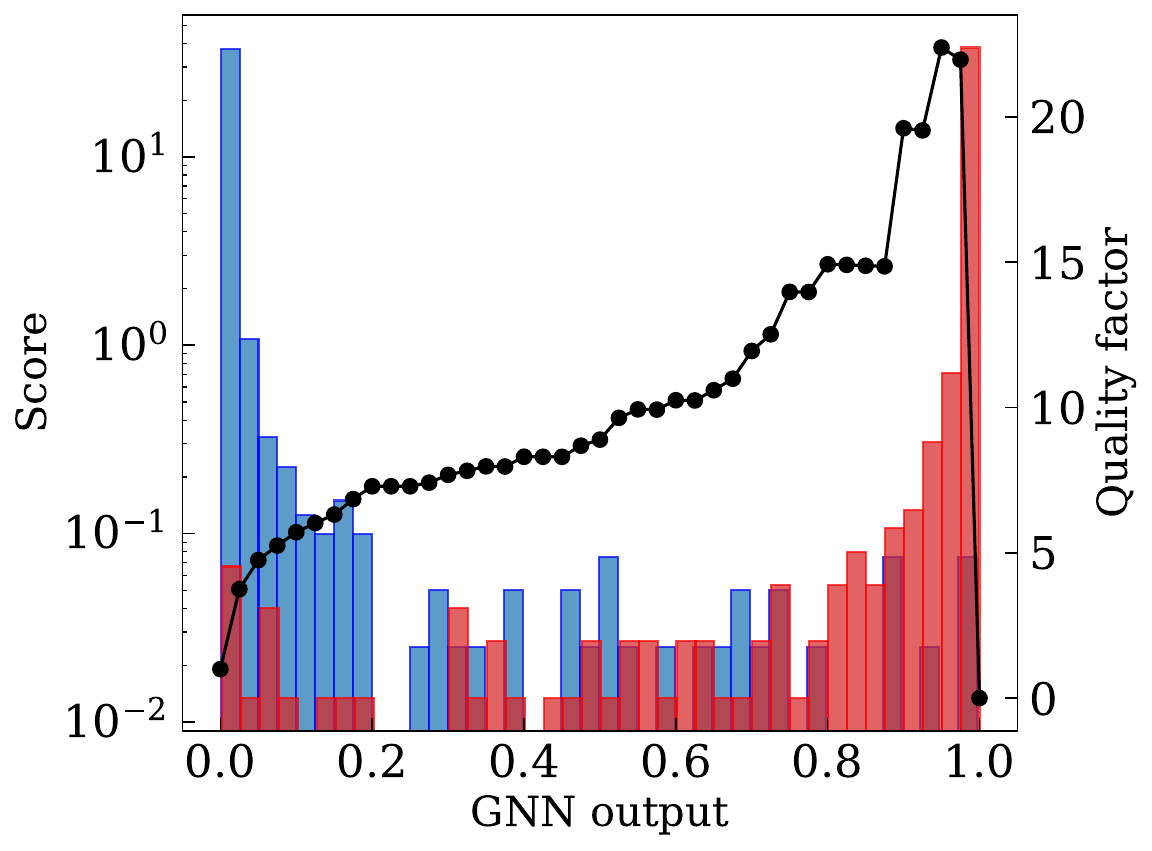}}\hfill
\caption{The GNN output for protons (blue) and gamma rays (red) using events in two different energy bins. The quality factor $q$ is shown in black here as a function of the score threshold (x-axis) of the GNN classifier. \label{fig:score_plot}}
\end{figure}

\subsubsection{Performance comparisons}
To compare the GNN to an established baseline, we investigated the background contamination for the different $\gamma$\;/\;hadron separators as a function of energy in Figure~\ref{fig:contamination_comparison_fixed_gammaeff}.
To enable a direct comparison, the gamma-ray efficiency was fixed to 80\%.
Note that we did not apply any quality cuts on the data.
Statistical uncertainties are estimated following Ref.~\cite{ullrich2012treatmenterrorsefficiencycalculations}.
If no background event is present in an energy bin, an upper limit is estimated by assuming a single background event and depicted by a down-facing arrow.
As can be seen, the GNN (red triangles) features a more efficient background rejection than LIC (green pentagons) and PINCness (purple stars) across the full studied range and even outperforms the combination of both observables (black triangles).
Whereas at low energies, the rejection is improved by around a factor of two, at higher energies, we find an improvement by one order of magnitude.

\begin{figure}[t!]
    \centering
    \includegraphics[width=0.9\textwidth]{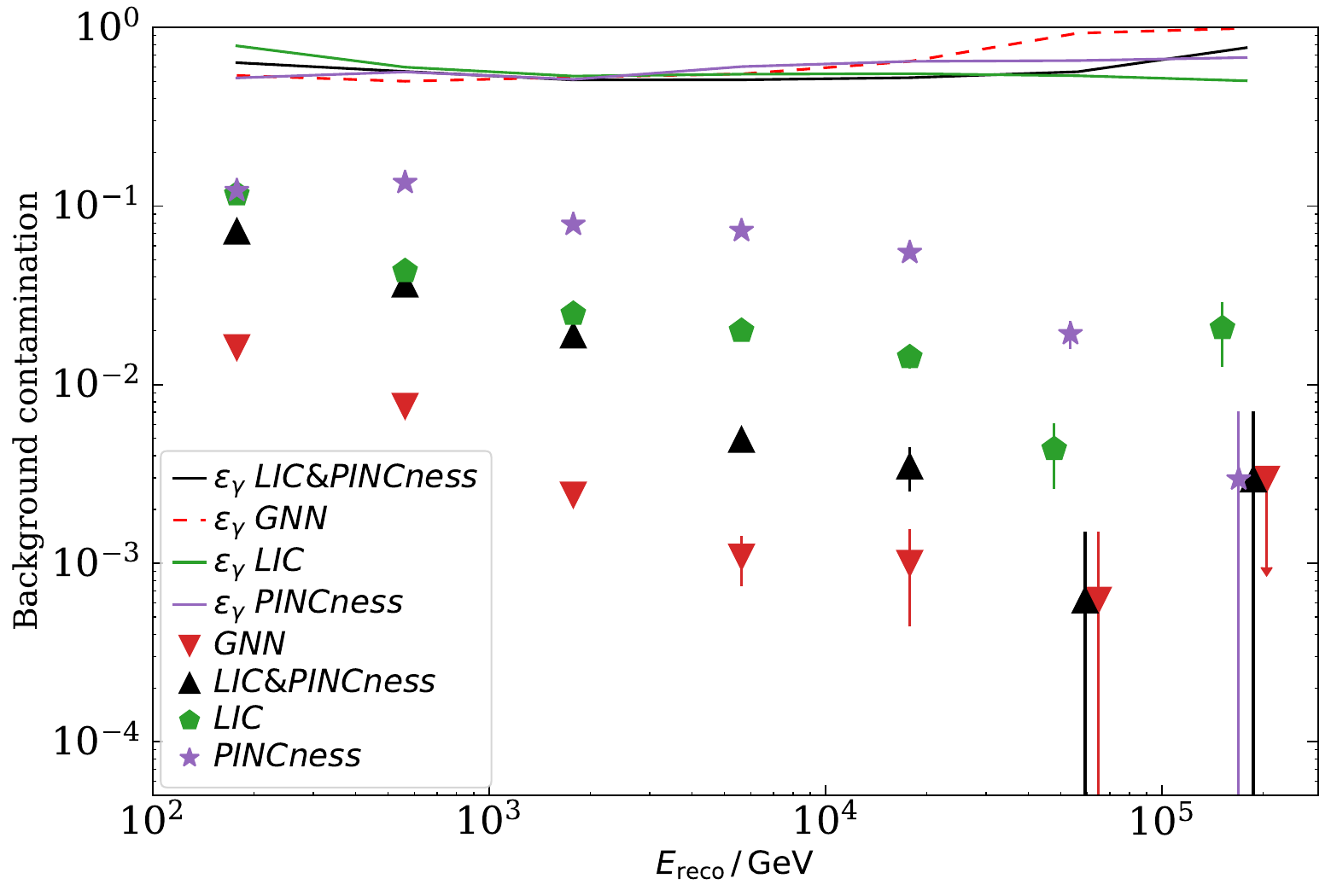}
    \caption{$\gamma$\;/\;hadron separation performance comparing the GNN to PINCness, LIC, and their combination without any quality cuts applied. The cuts were chosen to maximize the quality factor for a gamma-ray efficiency (shown as solid lines) of at least 50\%. The data points in the last two bins shifted horizontally to improve readability. Error bars denote statistical uncertainties. Downward-facing arrows indicate upper limits for bins if no background remains. No quality cuts have been applied.}
    \label{fig:contamination_comparison}
\end{figure}

\paragraph{Quality factor}
Since the quality of the $\gamma$\;/\;hadron separation directly propagates into the sensitivity that depends on $\epsilon_p$ and $\epsilon_\gamma$, the quality factor can be used to optimize source detections in the Gaussian limit.
In each bin, we optimized the quality factor (requiring a gamma-ray efficiency of at least 50\% to retain sufficient signal) for the different methods respectively.
Example plots for a low-energy and high-energy bin are shown in Figure~\ref{fig:score_plot} for the GNN, where the score distribution of the protons (blue histogram) and gamma rays (red histogram) are shown alongside the quality factor (black line).
In case there are no surviving background events, we instead estimate a lower limit of the quality factor with a single background event and indicate the limits with down-facing arrows.
To examine the effect of using the GNN on source sensitivity, we show in Figure~\ref{fig:contamination_comparison} the background contamination (colored markers) for the investigated approaches and the gamma-ray efficiency (colored lines) as a function of energy.
In terms of standalone parameters, LIC, shown in green, generally outperforms PINCness, shown in purple, in the low and medium energy range.
We find that our GNN, shown in red, outperforms the combination of LIC\,\&\,PINCness, depicted in black, by a factor of three at low energies and reaches a factor of five up to eight at medium and high energies.

\begin{figure}[t!]
    \centering
    \includegraphics[width=0.9\textwidth]{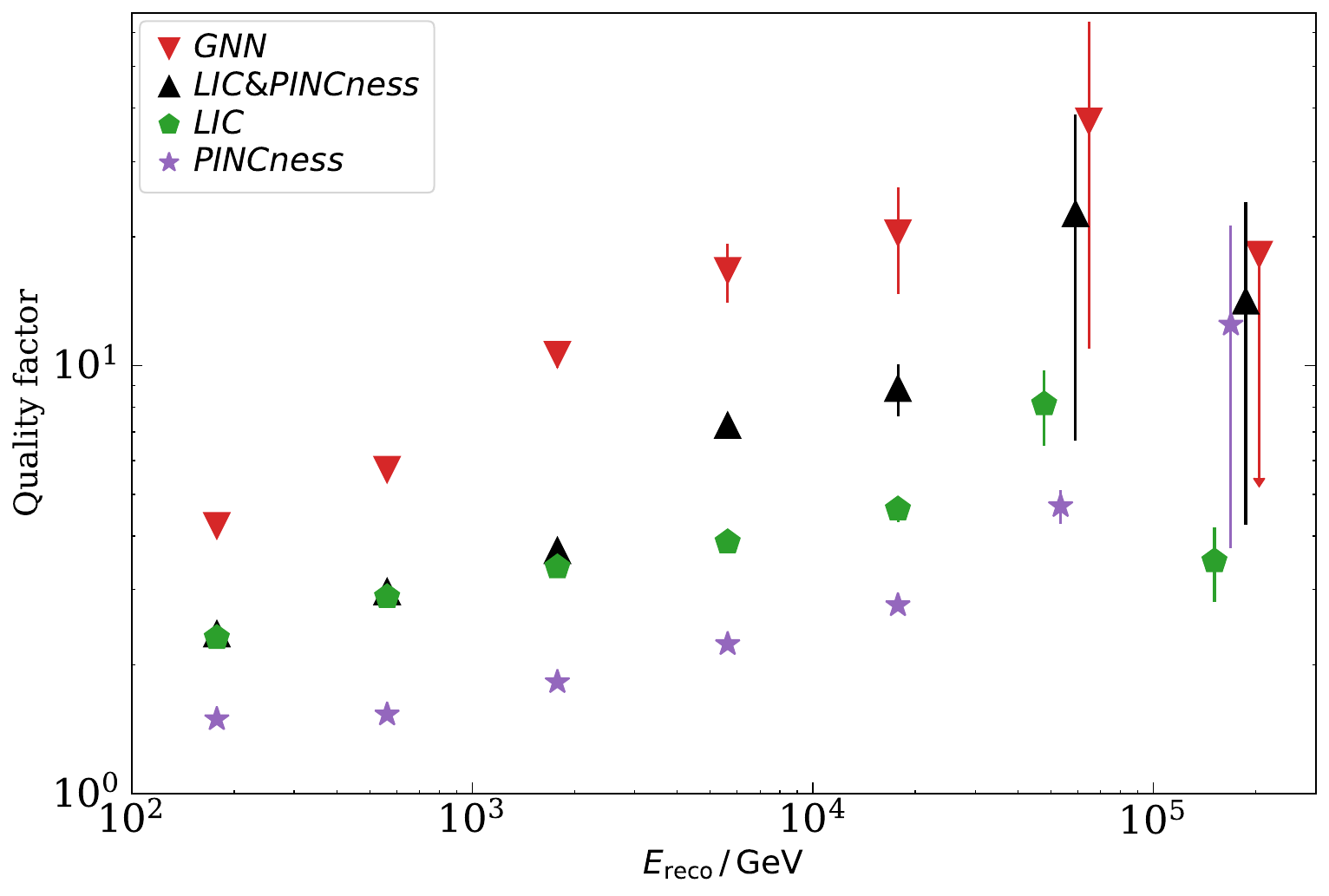}
    \caption{Comparison of the optimized quality factor for GNN (red), PINCness (purple), LIC (green), and their combination (black) without any quality cuts applied. Cuts were chosen to ensure at least 50\% gamma-ray efficiency. Upper limits are estimated assuming a single background event in the bin. Due to limited statistics resulting in large errors, the data points were shifted to avoid overlap.}
    \label{fig:qf_plot}
\end{figure}

The comparison of the optimized quality factor as a function of energy is shown in Figure~\ref{fig:qf_plot}.
The error bars denote statistical uncertainties and were estimated following Ref.~\cite{ullrich2012treatmenterrorsefficiencycalculations}.
For bins without background contamination, indicated by a down-facing array below the marker, a lower limit is estimated, assuming a single background event.
It can be seen that the GNN (red triangles) outperforms the classical observables (black triangles, green pentagons, and purple stars) over the whole energy range.
This demonstrates that the GNN efficiently extracts additional information from the sampled shower footprint, facilitating a significant gain in the signal-to-noise ratio.
Overall, the GNN offers a two to three times higher quality factor over the whole range and continues to improve the background rejection at very high energies.
However, the exact extent of the $\gamma$\;/\;hadron separation performance is somewhat unclear above 50\,TeV
given the limited statistics in these bins.

\section{Energy reconstruction using GNNs}
In general, we compare the performance of the energy reconstruction for events with zenith angle $\theta \leq 30\degree$ and $30\degree < \theta \leq 45\degree$. Showers at larger zenith angles have a much broader spatial spread across the array and have to travel a longer way through the atmosphere, which reduces the number of particles reaching the detector.\\
\paragraph{Data selection}
For the energy reconstruction, we apply a dedicated selection, which comprises cuts on the zenith angle $\theta \leq 45\degree$, the core position of the shower within the array ($r_\mathrm{array} < 300\,$m), and $n_{\text{hit}} > 30 $.
We also apply a cut on the score of the reconstructed events, which essentially acts as a quality cut for well-reconstructed gamma rays. The cut value was taken from the previous $\gamma$\;/\;hadron separation. In the energy reconstruction we only reconstruct gamma rays.

\subsection{Energy dispersion}
\begin{figure}[t!]
    \centering
    \subfloat[Zenith bin:  $\theta \leq 30 \degree$]{\includegraphics[width=0.49\textwidth]{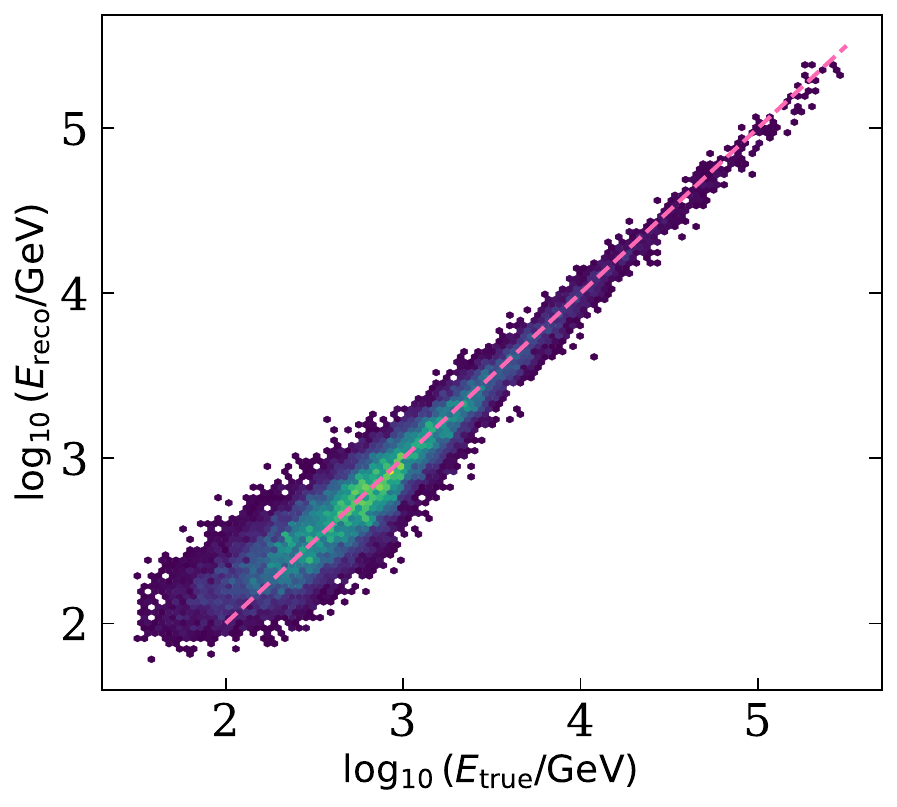}}
    \subfloat[Zenith bin:  $30 \degree < \theta \leq 45 \degree$]{\includegraphics[width=0.49
\textwidth]{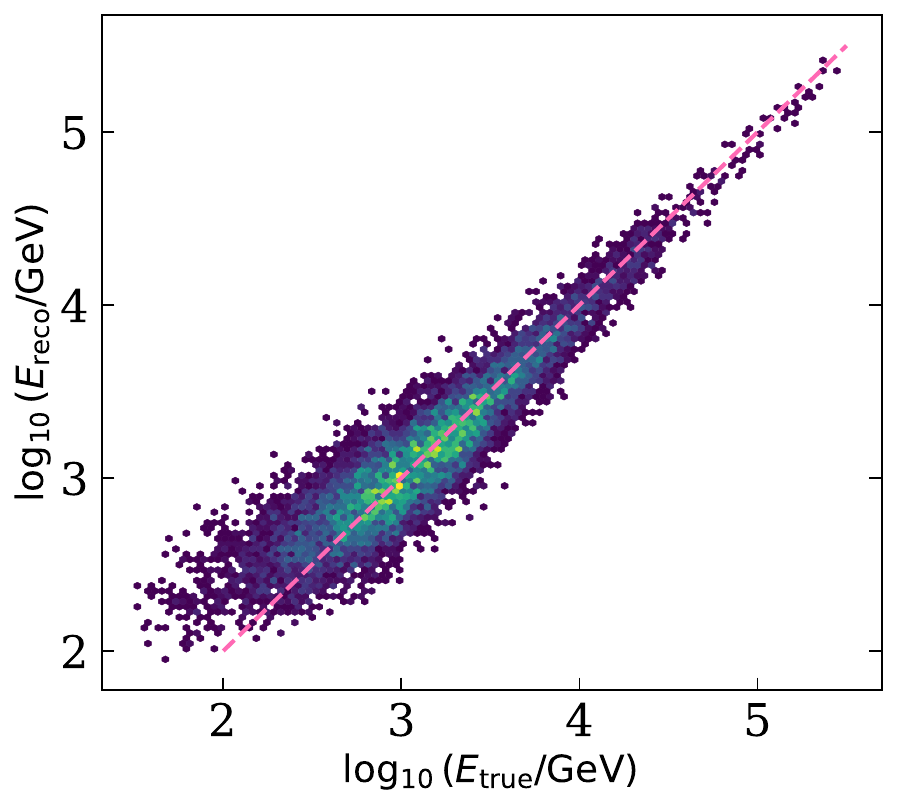}}\hfill
\caption{Energy dispersion matrix for two different zenith angle bins. The diagonal shows where $\log_{10}\left(E_{\text{reco}}\right) = \log_{10}\left(E_{\text{true}}\right)$.\label{fig:energy_dispertion}}
\end{figure}

In Figure~\ref{fig:energy_dispertion}, we compare the energy dispersion for events at smaller zenith angles Figure~\ref{fig:energy_dispertion}(a) with ones at larger zenith angles Figure~\ref{fig:energy_dispertion}(b).
We plot the reconstructed energy $\log_{10}\left(E_{\text{reco}}/{\mathrm{GeV}}\right)$ of the GNN versus its initial true energy $\log_{10}\left(E_{\text{true}}/{\mathrm{GeV}}\right)$. 
In an ideal case, the events in the plot align along the diagonal where $\log_{10}\left(E_{\text{reco}}/{\mathrm{GeV}}\right) = \log_{10}\left(E_{\text{true}}/{\mathrm{GeV}}\right)$. 
In both cases, we find that while there is a slight shift towards higher reconstructed energies, the distribution of events aligns well along this diagonal. 
Within each zenith bin, we find that the distribution at smaller energies deviates slightly more from the diagonal than at higher energies. 
This presumably stems from higher energy showers triggering more tanks, providing more information to the network.
Additionally, lower energies are more affected by upward fluctuations because they are close to the energy threshold of the detector.
We find that the overall distribution in the higher zenith angle bin is slightly broader than the one at lower zenith angles.
This is caused by the fact that the shower has to travel a longer distance through the atmosphere, leading to increased fluctuations in the shower development.
We find that there are more low-energy events at lower zenith angles than at higher angles.
Low-energy showers at high zenith angles are less likely to reach the array due to atmospheric absorption effects, thus changing the energy threshold.\\

\subsection{Energy bias and resolution}
To judge the performance of the GNN, we investigate the energy bias and energy resolution, which we define as the mean and standard deviation of the distribution $\log_{10}(E_{\text{reco}}/E_{\text{true}})$ and its uncertainties which we obtain by bootstrapping.
We compare the GNN results to the performance that can be attained using the current state-of-the-art in~\cite{templates_vikas} and~\cite{LHLatDistFit_PoS2023}, which relies on a template-based likelihood fitting procedure.\\

\begin{figure}[t!]
    \centering
    \includegraphics[width=0.95\textwidth]{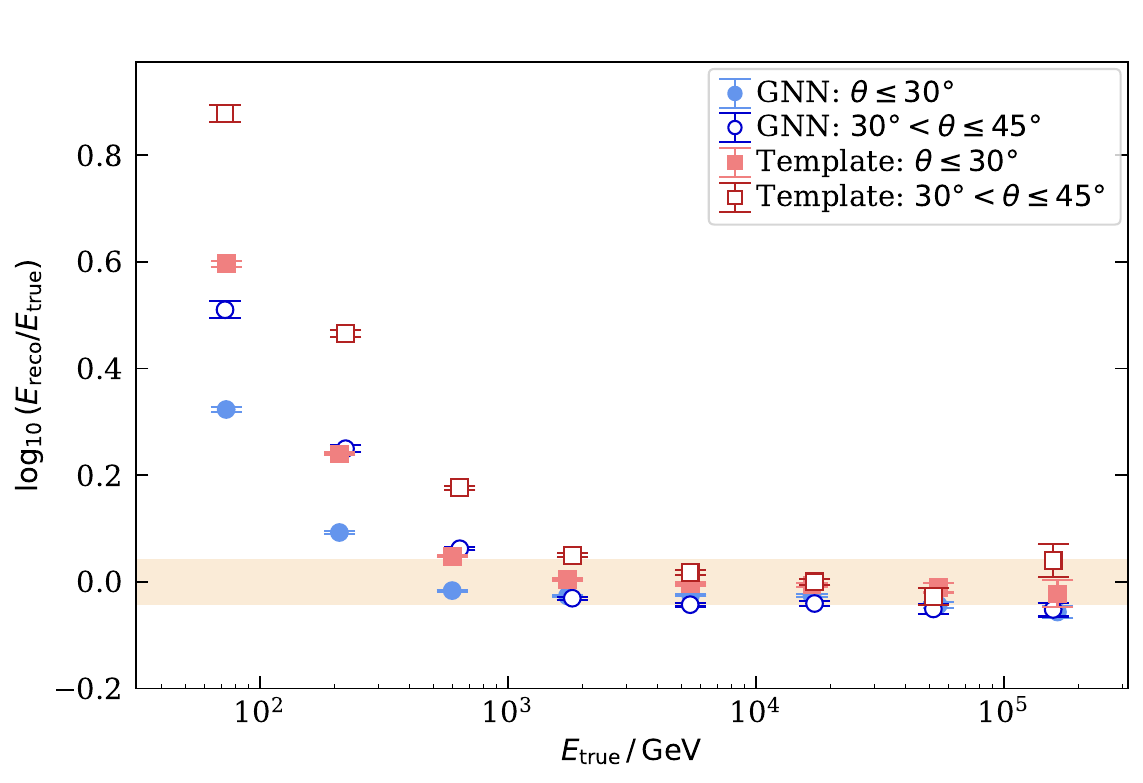}
    \caption{Comparison of the energy bias using the current state-of-the-art template approach~\cite{templates_vikas} and~\cite{LHLatDistFit_PoS2023}, which uses a likelihood fitting procedure (square markers) and the GNN (circles), as a function of energy for two zenith bins with filled markers for the $\theta \leq 30\degree$ bin and open markers for the $30\degree < \theta \leq 45\degree$ bin. 
    The bias is estimated as the mean of the distribution of $\log_{10}(E_{\rm{reco}}/E_{\rm{true}})$.
    The yellow marked area shows the $\pm 10\%$ region around 0 for the energy bias in linear.}
    \label{fig:energy_bias}
\end{figure}

\begin{figure}[t!]
    \centering
    \includegraphics[width=0.95\textwidth]{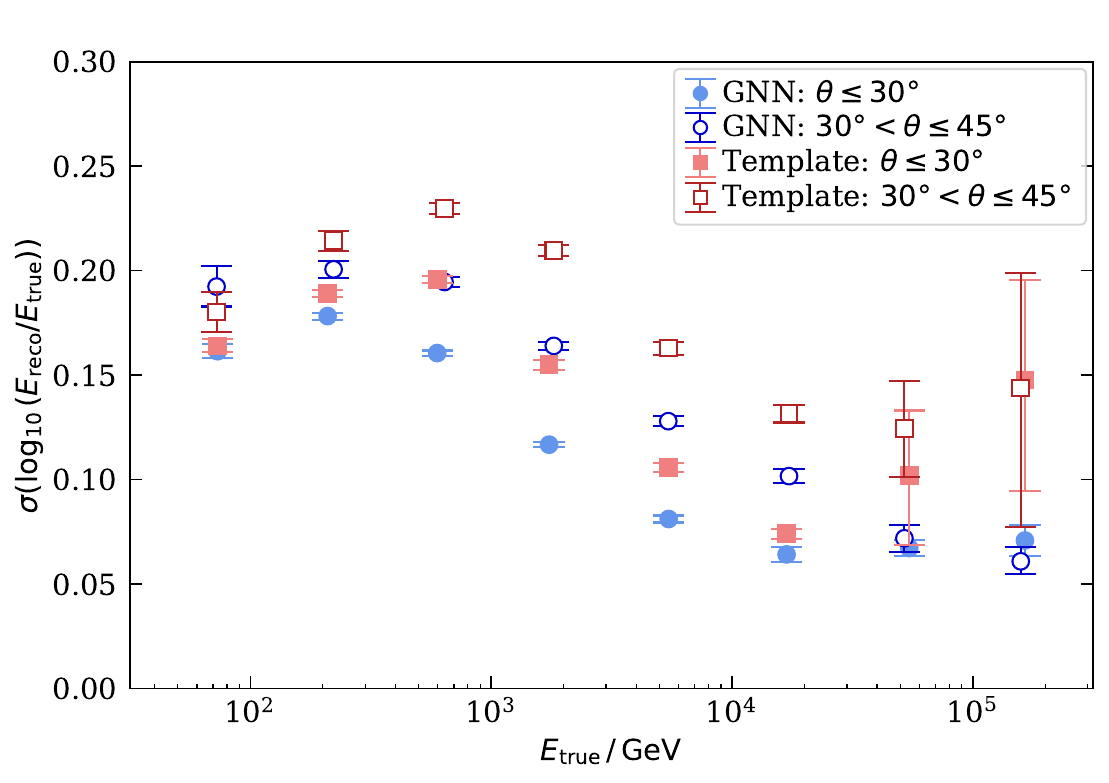}
    \caption{Comparison of the energy resolution using the the current state-of-the-art template approach~\cite{templates_vikas} and~\cite{LHLatDistFit_PoS2023}, which uses a likelihood fitting procedure (square markers) and the GNN (circles), as a function of energy for two zenith bins with filled markers for the $\theta \leq 30\degree$ bin and open markers for the $30\degree < \theta \leq 45\degree$ bin. We calculate the resolution as the standard deviation of the distribution of $\log_{10}(E_{\rm{reco}}/E_{\rm{true}})$.}
    \label{fig:energy_resolution}
\end{figure}

In Figure~\ref{fig:energy_bias}, the energy bias and in Figure~\ref{fig:energy_resolution}, the energy resolution is shown as a function of the energy. 
We show the results of the GNN-based energy reconstruction as circles compared to the performance that can be attained for the standard template method as squares for different zenith angle bins, filled markers for the  $\theta \leq 30\degree$ and open markers for $30\degree < \theta \leq 45\degree$.
The energy bias is stable over multiple orders of magnitude from 600\,GeV to around 200\,TeV.
Where the bias for both methods stays roughly within 10\% on a linear scale, this region is marked in yellow.
At lower energies below 600\,GeV, reconstruction is biased.
Since low-energetic events are less likely to be detected, only events with upwards fluctuations, i.e. showers that feature a shower maximum close to the detector, are measured.
This shifts the bias to higher energies, which affects the standard method more than the GNN.
The bias for the smaller zenith angle bin generally performs better compared to higher angles, which can again be explained by common shower physics.\\

In Figure~\ref{fig:energy_resolution}, we compare the energy resolution for the standard method as squares and the GNN method in circles for different zenith angle bins.
Again, we show filled markers for the  $\theta \leq 30\degree$ bin and open markers for the $30\degree < \theta \leq 45\degree$ bin.
For energies below 200\,GeV for the GNN, and below around 600\,GeV for the template method, we find that the energy resolution is getting slightly better with decreasing energy. 
The reason for this is that the cut on the number of tanks hit acts as a quality cut for events in that region.
As it is harder for low-energy showers to reach the threshold for reconstruction, more events are removed, artificially boosting the resolution in this region with the cost of a larger bias (compare the trend below 600~GeV in Figure~\ref{fig:energy_bias} and Figure~\ref{fig:energy_resolution}).
Above 600\,GeV, this threshold no longer influences the results.
Beyond 200\,GeV across the entire energy range, we find the GNN resolution is consistently surpassing what is currently achievable with the standard method.
At the highest energy above 100\,TeV, the resolution seems to worsen slightly, presumably due to the size of the shower footprint getting larger than the size of the array.\\

For the GNN resolution, we reach a resolution of around 16\% (linear scale) at around 17\,TeV for zenith angles $\theta \leq 30\degree$ and a resolution of around 26\% at the same energy for higher zenith angles.\\

Overall, we find a stable energy bias over many orders of magnitude for both the template method and the GNN and a very good energy resolution, especially medium energies, with the GNN showing evident improvements over the standard state-of-the-art method and improving over the performance requirement in the science proposal of SWGO~\cite{Albert:2019afb}.

\section{Conclusion}
In this work, we developed a novel deep-learning-based strategy for event reconstruction and $\gamma$\;/\;hadron separation in the context of SWGO, a next-generation gamma-ray observatory in the southern hemisphere currently in the research and development phase.
By modeling air shower footprints detected by a simulated water-Cherenkov detector array as graphs, we exploit the spatio-temporal patterns comprising the measured integrated signals and the timing information using graph neural networks (GNNs).
In the context of our simulation study, we examined the graph-based algorithms on energy reconstruction and $\gamma$\;/\;hadron separation for a detector design currently studied within the SWGO collaboration.

We presented the first comprehensive study of the application of graph neural networks to a water-Cherenkov-based gamma-ray observatory.
The obtained performance within this study surpasses the performance expected in the SWGO white paper~\cite{Albert:2019afb}.
We found that without the need for any quality cuts, the GNN provides a strong background rejection, outperforming hand-design variables currently in use within HAWC over the whole energy range.
These improvements are achieved by combining the spatial structure of the footprint, which is particularly efficient in discriminating showers at high energies, with the charge and timing information.
Interestingly, the timing information, which has not been investigated in detail for $\gamma$\;/\;hadron separation, appears to have additional separation power.

Furthermore, we found that GNNs offer an accurate energy reconstruction that provides reliable energy estimates from 600~GeV over the full studied energy range.
The GNN reconstruction offers improvements to the template-based approach~\cite{LHLatDistFit_PoS2023}, which is currently state-of-the-art, demonstrating an improved resolution, particularly at mid and high energies.
Due to the flexible nature of the proposed GNN architecture, changes in the layout and tank design, e.g., adding second chambers to enable the tagging of single muons, can be simply integrated into the presented algorithm.

Future work will focus on additional algorithm improvements, i.e., performing different clustering within the central and outer zones, utilizing attention mechanisms, and investigating its performance under realistic operation conditions, including cosmic ray noise.
This aims to exploit the full information contained in the detected air shower footprint, ultimately providing improved capabilities to survey the gamma-ray sky at very high energies.

\acknowledgments
We thank the SWGO Collaboration for allowing us to use SWGO simulations and to make use of the SWGO reconstruction software for this publication and the use of the common shared software framework (AERIE)~\cite{aerie}, kindly provided by HAWC.
We would like to particularly express our gratitude to R.~Turner and X.~Wang for their work on the PINCness observable in the SWGO reconstruction chain.
The authors gratefully acknowledge the scientific support and HPC resources provided by the Erlangen National High Performance Computing Center (NHR@FAU) of the Friedrich-Alexander-Universität Erlangen-Nürnberg (FAU) under the NHR project b129dc.
NHR funding is provided by federal and Bavarian state authorities.
NHR@FAU hardware is partially funded by the German Research Foundation (DFG) – 440719683.

\bibliographystyle{unsrtnat}
\bibliography{bibliography.bib}{}

\appendix

\newpage

\section{\label{sec:gnn_details}Network architectures}

\subsection{Parameters after hyperparameter search}
\begin{table}[h!]
    \centering
    \begin{centering}
            \begin{tabular}{ r r r r r}
                Parameters &  &$\gamma$\;/\;hadron separation & Energy reconstruction & range \\ 
                \hline\hline
                Learning rate & & 0.005 & 0.010 & [$5\text{e-}5$, $1.5\text{e-}2$]\\ 
                Decay factor &  & 0.39 & 0.29 & [0.2, 0.8]\\                 
                Batchsize &  & 135 & 105 & [96, 150]\\ 
                Weight decay &  & 0 (fixed) & $8.5\text{e-}4$ & [$5\text{e-}5$, $1\text{e-}3$] \\ 
                $n_\mathrm{EdgeConv}$ &  & 3 & 3 & [1, 4]\\ 
                $n_\mathrm{DynEdgeConv}$ &  & 1 & 1  & [1, 2]\\ 
                $n_\mathrm{feat}$ &  & 57 & 124 & [16, 128]\\ 
                $n_\mathrm{ResNet}$ &  & 5 & 3 & [1, 5] \\ 
                Batchnorm &  & False & True & [True, False] \\ 
                Dropout &  & 0.13 & 0.24 & [0, 0.5]\\ 
                $n_{k\mathrm{NN, DynEdgeConv}}$ &  & 16 (fixed) & 13 & [8, 16]\\ 
                
            \end{tabular}
            \caption{Found parameters after the hyperparameter search for $\gamma$\;/\;hadron separation and energy reconstruction. $n_\mathrm{EdgeConv}$ and $n_\mathrm{DynEdgeConv}$ denote the number of EdgeConvolution and DynamicEdgeConvolution layers, and $n_\mathrm{resnet}$ the number of ResNet blocks.
            $n_{\mathrm{feat}}$ is the number of features and $n_{k\mathrm{NN, DynEdgeConv}}$ the number of $k$NN neighbors for the DynamicEdgeConvolution.}
            \label{tab:edge_conv_arch}
    \end{centering}
\end{table}

\subsection{Task specific architectures}
\FloatBarrier

\begin{table}[h!]
    \centering
    \begin{centering}
            \begin{tabular}{ r c r r r r }
                            \multicolumn{6}{l}{Architecture for $\gamma$\;/\;hadron separation} \\
                \hline\hline
                 & Layer & features & setting & input shape & output shape \\ 
                \hline
                 & EdgeConv & $n_\mathrm{feat}$ & $\aggfn_{j}$: mean & $n_\mathrm{nodes} \times 4$ & $n_\mathrm{nodes} \times n_\mathrm{feat}$ \\
                 & ReLU & -- & -- & $n_\mathrm{nodes} \times n_\mathrm{feat}$ & $n_\mathrm{nodes} \times n_\mathrm{feat}$ \\ 
                 \multirow{2}{*}{$3 \times \Big{\{}$} & EdgeConv & $n_\mathrm{feat}$ & $\aggfn_{j}$: mean & $n_\mathrm{nodes} \times n_\mathrm{feat}$ & $n_\mathrm{nodes} \times n_\mathrm{feat}$ \\ 
                 & ReLU & -- & -- & $n_\mathrm{nodes} \times n_\mathrm{feat}$ & $n_\mathrm{nodes} \times n_\mathrm{feat}$ \\ 
                 & DynEdgeConv & $n_\mathrm{feat}$ & $\aggfn_{j}$: mean & $n_\mathrm{nodes} \times n_\mathrm{feat}$ & $n_\mathrm{nodes} \times n_\mathrm{feat}$ \\
                 & Concat & -- & layer outputs & $ 4 \cdot (n_\mathrm{nodes} \times n_\mathrm{feat})$ & $ 4 \cdot (n_\mathrm{nodes} \times n_\mathrm{feat})$ \\
                 & MaxPooling & -- & -- & $ 4 \cdot (n_\mathrm{nodes} \times n_\mathrm{feat})$ & $ 4 \cdot (n_\mathrm{nodes} \times n_\mathrm{feat})$\\
                 & Dropout & -- & $p=0.13$ & $ 4 \cdot (n_\mathrm{nodes} \times n_\mathrm{feat})$ & $ 4 \cdot (n_\mathrm{nodes} \times n_\mathrm{feat})$ \\
                & ResNet & $n_{\mathrm{feat}} $ & -- & $ 4 \cdot (n_\mathrm{nodes} \times n_\mathrm{feat})$ & $n_{\mathrm{feat}}$ \\
                 $4 \times \{$ & ResNet & $n_{\mathrm{feat}} $ & -- & $n_{\mathrm{feat}}$ & $n_{\mathrm{feat}}$ \\
                 & Dropout & -- & $p=0.13$ & $n_{\mathrm{feat}}$ & $n_{\mathrm{feat}}$ \\
                 & Linear & 2 & -- & $n_{\mathrm{feat}}$ & $2$ \\
                 & SoftMax & -- & -- & $n_{\mathrm{feat}}$ & $2$ \\

                \hline
            \end{tabular}
            \caption{Network architectures for the separation task. 
            $n_\mathrm{nodes}$ describes the number of nodes in the input graph, which depend on the number of triggered tanks, and $n_{\mathrm{feat}}$ the number of features.
            We further use a bottleneck layer in the first ResNet layer to adjust the dimensionality.}
            \label{tab:separation_arch}
    \end{centering}
\end{table}

\begin{table}[h!]
    \centering
    \begin{centering}
            \begin{tabular}{ r c r r r r }
                            \multicolumn{6}{l}{Architecture for energy reconstruction} \\
                \hline\hline
                 & Layer & features & setting & input shape & output shape \\ 
                \hline
                 & EdgeConv & $n_\mathrm{feat}$ & $\aggfn_{j}$: mean & $n_\mathrm{nodes} \times 4$ & $n_\mathrm{nodes} \times n_\mathrm{feat}$ \\ 
                 & Linear & $n_\mathrm{feat}$ & -- & $n_\mathrm{nodes} \times 4$ & $n_\mathrm{nodes} \times n_\mathrm{feat}$\\ 
                 & Addition & -- & -- & $n_\mathrm{nodes} \times 4$ & $n_\mathrm{nodes} \times n_\mathrm{feat}$ \\
                 & ReLU & -- & -- & $n_\mathrm{nodes} \times n_\mathrm{feat}$ & $n_\mathrm{nodes} \times n_\mathrm{feat}$ \\ 
                 \multirow{4}{*}{$3 \times \left \{ \rule{0em}{9.5mm} \right.$} & EdgeConv & $n_\mathrm{feat}$ & $\aggfn_{j}$: mean & $n_\mathrm{nodes} \times n_\mathrm{feat}$ & $n_\mathrm{nodes} \times n_\mathrm{feat}$ \\ 
                 & Linear & $n_\mathrm{feat}$ & -- & $n_\mathrm{nodes} \times n_\mathrm{feat}$ & $n_\mathrm{nodes} \times n_\mathrm{feat}$\\ 
                 & Addition & -- & -- & $n_\mathrm{nodes} \times n_\mathrm{feat}$ & $n_\mathrm{nodes} \times n_\mathrm{feat}$ \\
                  & ReLU & -- & -- & $n_\mathrm{nodes} \times n_\mathrm{feat}$ & $n_\mathrm{nodes} \times n_\mathrm{feat}$ \\ 
                 & DynEdgeConv & $n_\mathrm{feat}$ & $\aggfn_{j}$: mean & $n_\mathrm{nodes} \times n_\mathrm{feat}$ & $n_\mathrm{nodes} \times n_\mathrm{feat}\cdot 2$ \\ 
                 & Linear & $n_\mathrm{feat}$ & -- & $n_\mathrm{nodes} \times n_\mathrm{feat}$ & $n_\mathrm{nodes} \times n_\mathrm{feat} \cdot 2$\\ 
                 & Addition & -- & -- & $n_\mathrm{nodes} \times n_\mathrm{feat}$ & $n_\mathrm{nodes} \times n_\mathrm{feat}\cdot 2$ \\ 
                 & ReLU & -- & -- & $n_\mathrm{nodes} \times n_\mathrm{feat}\cdot 2$ & $n_\mathrm{nodes} \times n_\mathrm{feat}\cdot 2$ \\ 
                 & Concat & -- & layer outputs & $ 5 \cdot (n_\mathrm{nodes} \times n_\mathrm{feat})$ & $ 5 \cdot (n_\mathrm{nodes} \times n_\mathrm{feat})$ \\
                 & MaxPooling & -- & -- & $ 5 \cdot (n_\mathrm{nodes} \times n_\mathrm{feat})$ & $ 5 \cdot (n_\mathrm{nodes} \times n_\mathrm{feat})$\\
                 & Batchnorm & -- & momentum\,=\,0.1 & $ 5 \cdot (n_\mathrm{nodes} \times n_\mathrm{feat})$ & $ 5 \cdot (n_\mathrm{nodes} \times n_\mathrm{feat})$\\
                 & Dropout & -- & $p=0.24$ & $ 5 \cdot (n_\mathrm{nodes} \times n_\mathrm{feat})$ & $ 5 \cdot (n_\mathrm{nodes} \times n_\mathrm{feat})$ \\
                & ResNet & $n_{\mathrm{feat}} $ & -- & $ 5 \cdot (n_\mathrm{nodes} \times n_\mathrm{feat})$ & $n_{\mathrm{feat}}$ \\
                 $2 \times \{$ & ResNet & $n_{\mathrm{feat}} $ & -- & $n_{\mathrm{feat}}$ & $n_{\mathrm{feat}}$ \\
                 & Dropout & -- & $p=0.24$ & $n_{\mathrm{feat}}$ & $n_{\mathrm{feat}}$ \\
                 & Linear & 1 & -- & $n_{\mathrm{feat}}$ & $1$ \\

                \hline
            \end{tabular}
            \caption{Network architectures for the energy reconstruction task. 
            $n_\mathrm{nodes}$ describes the number of nodes in the input graph, which depend on the number of triggered tanks, and $n_{\mathrm{feat}}$ the number of features. 
            The ``Addition'' denotes a residual connection that connects consecutive convolutional and linear layers.
            We further use a bottleneck layer in the first ResNet layer to adjust the dimensionality.}
            \label{tab:energy_reco_arch}
    \end{centering}
\end{table}

\FloatBarrier

\subsection{Kernel function and residual model}

\begin{table}[bh!]
    \begin{centering}
        \begin{subtable}[h]{0.495\textwidth}
            \begin{tabular}{ c r r }
                \multicolumn{3}{l}{Kernel network $h_\mathbf{\Theta}$} \\
                \hline\hline
                layer & features & settings \\
                \hline
                Linear & $n_\mathrm{feat}$ & no bias \\
                Batchnorm & -- & momentum=0.9  \\
                Activation & --  & --  \\
                Linear & $n_\mathrm{feat}$ & no bias \\
                Batchnorm & -- & momentum=0.9  \\
                Activation & -- & -- \\
                Linear & $n_\mathrm{feat}$ & no bias \\
                Batchnorm & -- & momentum=0.9  \\
                Activation & -- & -- \\
                \hline
            \end{tabular}
            \caption{Architecture of the kernel function $h_\mathbf{\Theta}$, where $n_\mathrm{feat}$ is to be chosen by the user. 
            For $\gamma$\;/\;hadron separation, we used SiLU as an activation function and for the energy reconstruction ReLU.
            }
            \label{tab:edge_conv_ffn}
        \end{subtable}
        \hfill
        \begin{subtable}[h]{0.495\textwidth}
            \begin{centering}
                \begin{tabular}{ c r r }
                    \multicolumn{3}{l}{Residual Module "ResNet"}\\
                    \hline\hline
                    layer & features & settings \\
                    \hline
                    Linear & $n_\mathrm{feat}$ & no bias \\
                    SiLU & --  & --  \\
                    Batchnorm & -- & momentum=0.1 \\
                    Linear & $n_\mathrm{feat}$ & no bias \\
                    SiLU & --  & --  \\
                    Batchnorm & -- & momentum=0.1 \\
                    Add & -- & -- \\
                    SiLU & -- & -- \\
                    \hline
                \end{tabular}
                \caption{Details of our used ResNet modules.}
                \label{tab:resnet_arch}
           \end{centering}
        \end{subtable}
    \end{centering}
    \label{tab:edge_conv_kernel}
\end{table}

\FloatBarrier

\newpage

\subsection{Validation losses}
\begin{figure}[h!]
    \centering
    \includegraphics[width=0.95\textwidth]{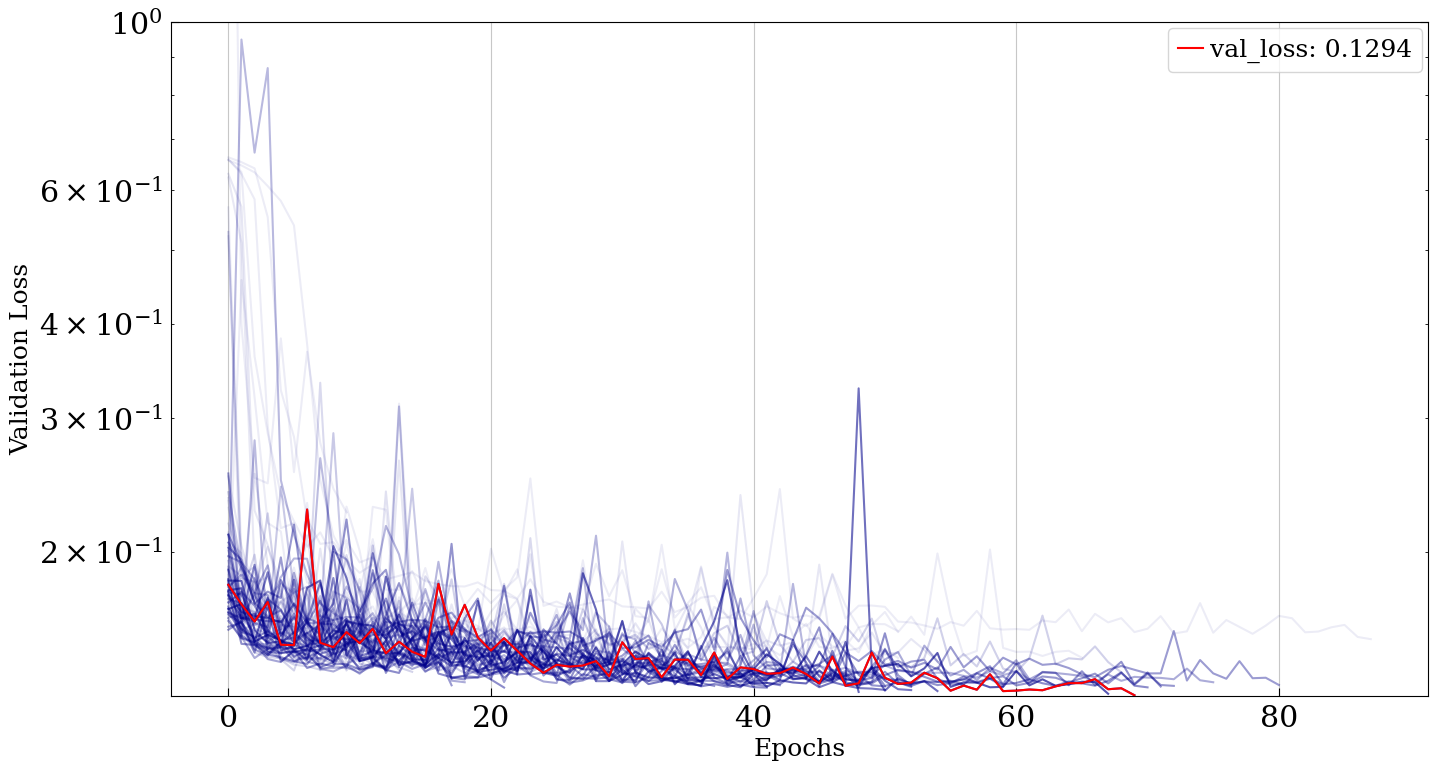}
    \caption{History of the validation losses during the hyperparameter search for $\gamma$\;/\;hadron separation. The best-performing model is shown in red.}
    \label{fig:gh_val_loss}
\end{figure}

\begin{figure}[h!]
    \centering
    \includegraphics[width=0.95\textwidth]{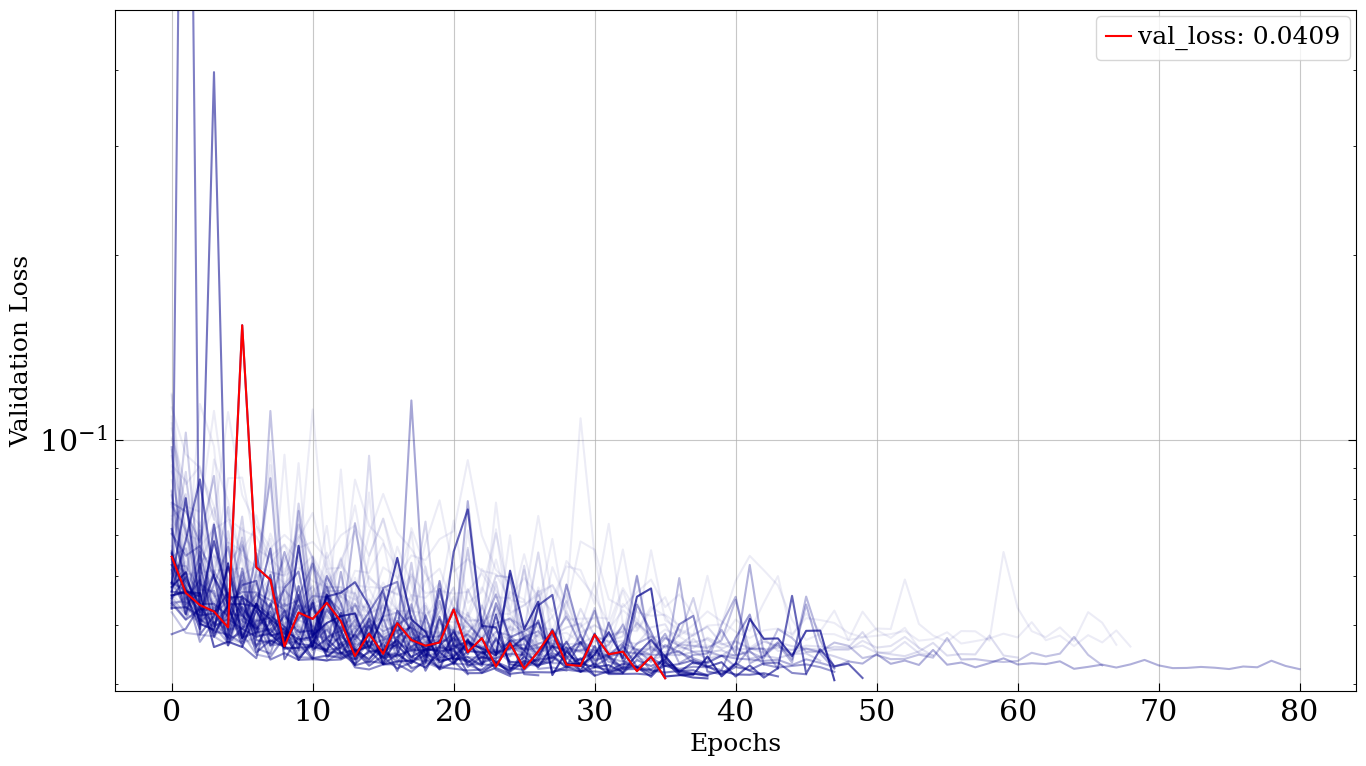}
    \caption{History of the validation losses during the hyperparameter search for energy reconstruction. The best-performing model is shown in red.}
    \label{fig:en_val_loss}
\end{figure}

\end{document}